\documentclass[fleqn,10pt]{jack}
\usepackage[utf8]{inputenc}
\usepackage[T1]{fontenc}
\usepackage{bm}
\usepackage{comment}
\usepackage{upgreek}
\usepackage[font=small]{caption}
\usepackage{amsmath}
\title{Reconfigurable Training and Reservoir Computing in an Artificial Spin-Vortex Ice via Spin-Wave Fingerprinting}

\author[1,$\dagger$,*]{Jack C. Gartside}
\author[1,$\dagger$]{Kilian D. Stenning}
\author[1,$\dagger$]{Alex Vanstone}
\author[1]{Holly H. Holder}
\author[3,4]{Daan M. Arroo}
\author[5,2]{Troy Dion}
\author[6]{Francesco Caravelli}
\author[2]{Hidekazu Kurebayashi}
\author[1,4]{Will R. Branford}
\affil[1]{Blackett Laboratory, Imperial College London, London SW7 2AZ, United Kingdom}
\affil[2]{London Centre for Nanotechnology, University College London, London WC1H 0AH, United Kingdom}
\affil[3]{Department of Materials, Imperial College London, London SW7 2AZ, United Kingdom}
\affil[4]{London Centre for Nanotechnology, Imperial College London, London SW7 2AZ, United Kingdom}
\affil[5]{Solid State Physics Lab., Kyushu University, 744 Motooka, Nishi-ku, Fukuoka, 819-0395, Japan}
\affil[6]{Theoretical Division (T4), Los Alamos National Laboratory, Los Alamos, New Mexico 87545, USA}
\affil[$\dagger$]{These authors contributed equally}
\affil[*]{Corresponding author e-mail: j.carter-gartside13@imperial.ac.uk}

\begin{abstract}
Strongly-interacting artificial spin systems are moving beyond mimicking naturally-occurring materials to emerge as versatile functional platforms, from reconfigurable magnonics to neuromorphic computing. 

Typically artificial spin systems comprise nanomagnets with a single magnetisation texture: collinear macrospins or chiral vortices. By tuning nanoarray dimensions we achieve macrospin/vortex bistability and demonstrate a four-state metamaterial spin-system `Artificial Spin-Vortex Ice’ (ASVI). ASVI can host Ising-like macrospins with strong ice-like vertex interactions, and weakly-coupled vortices with low stray dipolar-field. Vortices and macrospins exhibit starkly-differing spin-wave spectra with analogue mode-amplitude control and mode-frequency shifts of $\Delta f = 3.8~$ GHz.

The enhanced bi-textural microstate space gives rise to emergent physical memory phenomena, with ratchet-like vortex injection and history-dependent nonlinear fading memory when driven through global magnetic field cycles. We employ spin-wave microstate fingerprinting for rapid, scaleable readout of vortex and macrospin populations and leverage this for spin-wave reservoir computation. ASVI performs non-linear mapping transformations of diverse input and target signals in addition to chaotic time-series forecasting.
\end{abstract}

\begin{document}

\flushbottom
\maketitle
\thispagestyle{empty}


Artificial spin systems\cite{skjaervo2020advances} are metamaterials representing magnetic spins via magnetisation textures of nanomagnetic elements. Typically, nanoelement dimensions are tuned to energetically favour a single texture throughout the system, i.e. Ising-like macrospins in nanoislands\cite{wang2006artificial} or vortex-states in nanodisks\cite{shinjo2000magnetic}. These simple systems have enabled observation of diverse fundamental phenomena including emergent `magnetic-monopole' defects\cite{ladak2010direct} and spontaneous long-range ordering\cite{morgan2011thermal}. However, limiting systems to single textures places arbitrary constraints on the complexity of emergent behaviours\cite{yu2021magnetic}. The great freedom of artificial spin systems \& metamaterials at large is that properties may be tailored through nanofabrication, allowing complex `designer' behaviours not observed in nature. Recently, four-state systems were demonstrated using pairs of Ising-like macrospins to represent `metaspins',\cite{sklenar2019field} or square nanomagnets with four-state Potts behaviour\cite{louis2018tunable}, with enhanced microstate-spaces and corresponding emergent behaviours not observed in two-state Ising or vortex systems. These systems employ single magnetisation textures, expanding the texture range offers promising future directions.

Reconfigurable magnonics\cite{grundler2015reconfigurable,chumak2017magnonic,barman2020magnetization,kaffash2021nanomagnonics,barman20212021,papp2021nanoscale,dion2021observation,arroo2019sculpting,dion2019tunable,stenning2020magnonic,vanstone2021spectral,chaurasiya2021comparison}, physical memory\cite{keim2019memory} and hardware neuromorphic computation\cite{tanaka2019recent,nakajima2020physical,markovic2020physics,milano2021materia,chumak2021roadmap,papp2021nanoscale} are critical future technologies reliant on rich, diverse microstate-spaces to fulfill their promise. Reservoir computing (RC) schemes are one neuromorphic approach, competitive with deep neural-networks (DNN) across applications such as learning dynamic linear and nonlinear transformations, as well as time-series forecasting, at a fraction of DNN training time and energy cost\cite{tanaka2019recent,nakajima2020physical,milano2021materia,dawidek2021dynamically,torrejon2017neuromorphic}. The reservoir acts as a `black box' comprising randomly connected nonlinear elements, with training only performed on the system readout, offering huge energy-savings versus training every network weight as in DNNs. Spin-wave\cite{torrejon2017neuromorphic,nakane2018reservoir,chumak2021roadmap} and artificial spin\cite{hon2021numerical, jensen2018computation,jensen2020reservoir,welbourne2021voltage} reservoirs offer great processing potential, yet experimental realisation and scalability are often hampered by complex electrical connectivity requirements and inter-element coupling issues as well as lack of rapid, scaleable readout.

By tailoring nanoelements such that Ising and vortex states are energetically equivalent, we present artificial spin vortex ice (ASVI) - a four-state, bi-textured spin-system; two Ising-like macrospin orientations and two vortex chiralities. We define ASVI as a strongly-interacting nanomagnetic array. The strong interactions between Ising-like macrospins favour ice-rule configurations, whereas vortices exhibit flux-closure patterns with low stray dipolar-field. 

\begin{figure}[tbp]
\centering
\includegraphics[width=0.99\textwidth]{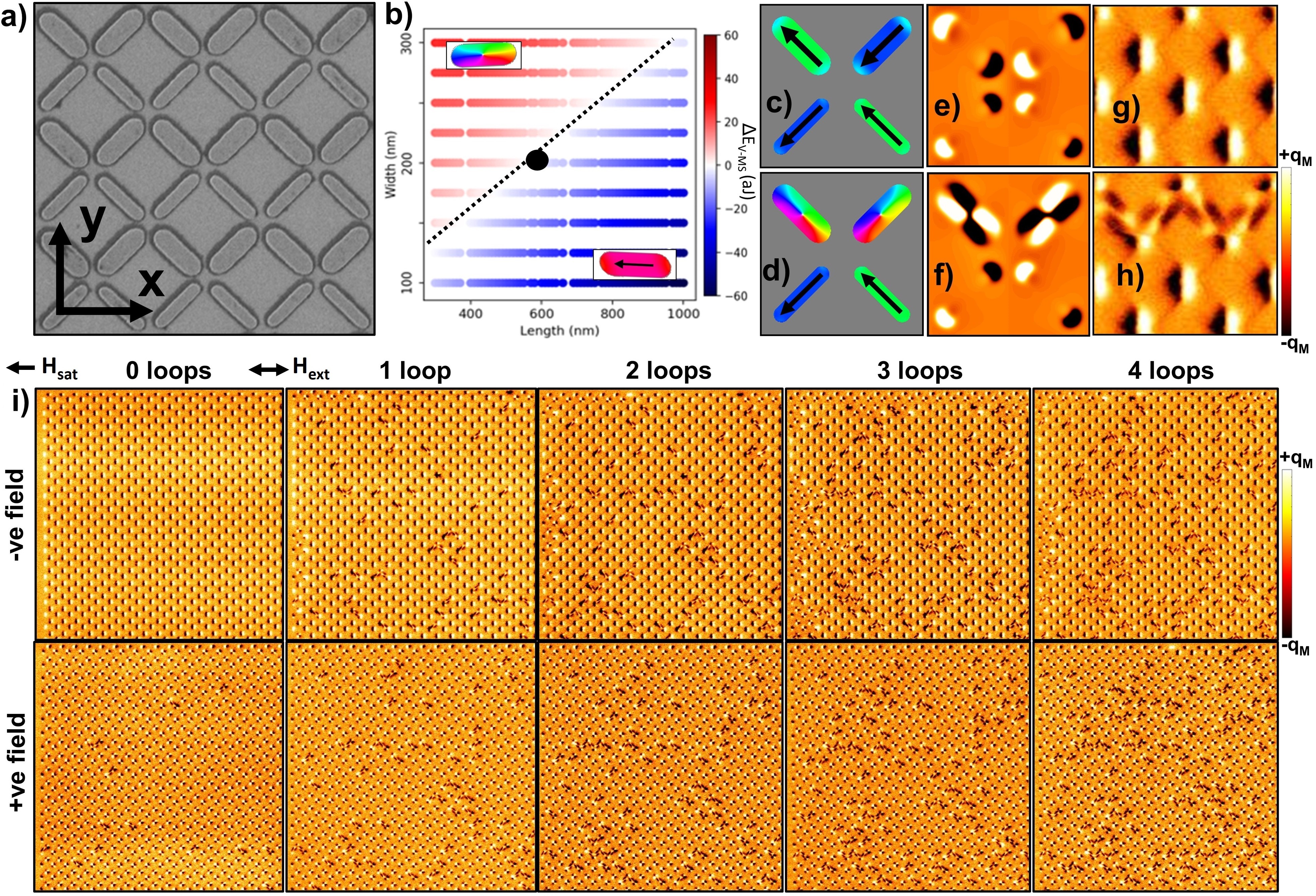}
\caption{Artificial spin-vortex ice.\\
a) Scanning electron micrograph (3.5 $\times$ 3.5 $\upmu$m$^2$) of artificial spin-vortex ice. Permalloy bars are 600 nm long, 200 nm (wide-bar) and 125 nm (thin-bar) wide, 20 nm thick with 100 nm vertex gap (bar-end to vertex-centre). Lattice vectors $\hat{x}$ (along wide/thin bar rows) and $\hat{y}$ (perpendicular to rows) are defined.\\
b) Phase diagram of energy-difference between vortex and macrospin states for a range of bar dimensions, determined via MuMax3 simulation. Red regions favour vortex states, blue favour macrospin. Dotted black line indicates dimensions with equal macrospin and vortex energy, black circle at 600 nm $\times$ 200 nm indicates wide-bar dimensions of sample studied here. \\
c-f) MuMax3-simulated magnetisation states of a single 4-bar vertex in all-macrospin (c) and wide-bar vortex, thin-bar macrospin (d) states. Simulated MFM images produced from magnetisation states (c,d), corresponding to (e,f) respectively. Images are 1.6 $\times$ 1.6 $\upmu$m$^2$. MFM signal is simulated for 60 nm tip height above sample. Vertex gap is exaggerated relative to the experimental sample to allow clearer visualisation.\\
g,h) Experimental magnetic force microscope images (1.9 $\times$ 1.9 $\upmu$m$^2$).  of all-macrospin (g) and mixed vortex-macrospin (h) states. h) shows vortices in the top 4 bars, macrospins in the lower 8. MFM scale bars are converted from phase shift to magnetic charge $q_M$.\\
i) Magnetic force micrograph series (15x15 $\upmu$m$^2$ images) showing four-loop vortex-injection field-sequence from initial $-\hat{x}$ saturated, pure-macrospin state (0 loop, -ve field). Field-loop amplitude is 18 mT, applied along $\hat{x}$. Series continues for 5, 7 and 10 loops, shown in supplementary figure 2. Formation of vortex and macrospin domains is observed as field-looping progresses to higher loop numbers. All images taken in zero-field.}
\label{Fig1}
\end{figure}

The bistable texture drives emergent physical-memory properties. Vortices are stable to higher magnetic field than macrospins (all references to `field' in this work refer to magnetic field), allowing ratchet-like vortex population control. Low vortex stray-field modifies the dipolar field landscape, pinning or promoting reversal of adjacent bars. This is leveraged for local control of memory and switching dynamics via complex, nonlinear, vortex population control protocols. The array incorporates thinner bars, tuned for macrospin-stability, providing a reconfigurable bias-field which exerts control over vortex-injection dynamics.

Magnons are highly-sensitive to magnetic texture\cite{yu2021magnetic} and the bistable ASVI textures offer deep magnonic reconfigurability, with $\Delta f = 3.8~$ GHz between vortex and macrospin modes and highly-nonlinear vortex field-gradients. We demonstrate fine analogue tuning of mode amplitudes via vortex-injection, affording exceptional levels of spectral control and reconfigurability relative to existing reconfigurable magnonic crystals. We employ spin-wave microstate fingerprinting\cite{vanstone2021spectral} for vortex/macrospin population readout and rapid, scaleable measurement of physical memory effects. The non-volatility of nanomagnetic states combined with history dependent dynamics results in a 'fading memory'\cite{yildiz2012re} i.e. the system gradually `forgets' prior inputs when exposed to new data (here via global field loops), a key requirement for RC. We utilise the physical memory and spin-wave properties of ASVI to realise a spin-wave reservoir computation scheme free from the need to electrically-address individual reservoir elements. We demonstrate learning of linear and non-linear signal transformations and chaotic time-series forecasting. We observe strong computational performance when employing length-constrained training datasets (here 200-400 points, 500-900 is typical\cite{milano2021materia,moon2019temporal}) to reflect real-world, device-based applications such as smartphones, space-exploration and internet of things (IoT) with strict constraints on battery life and data-capture.

\section*{Artificial spin-vortex ice}


The ASVI studied here is based on square lattice ASI\cite{morgan2011thermal}, with alternating rows of thin and wide bars along $\hat{x}$ (Figure \ref{Fig1} a). Different coercive fields for thin and wide bars permit global-field microstate control, as described previously\cite{gartside2021reconfigurable}. Bars are permalloy, 600 nm long, 200 nm (wide-bar) and 125 nm (thin-bar) wide and 20 nm thick with 100 nm vertex gap (bar-end to vertex-centre). Along $\hat{x}$, wide-bar coercive field distribution is $H_{\rm c1} = 15.5-17~$ mT, thin bar coercive field $H_{\rm c2} = 26-29$ mT. Wide-bar dimensions are chosen such that the combined demagnetisation and exchange energy of the macrospin state are equal to the vortex state (determined via micromagnetic simulation, Figure \ref{Fig1} b), giving macrospin and vortex bistability\cite{metlov2002stability,guslienko2008magnetic,talapatra2020magnetic}. The ASVI considered here is bicomponent, with macrospin/vortex transition occurring in wide-bars. Thin bars remain in macrospin states to provide a reconfigurable dipolar bias-field landscape. Figure \ref{Fig1} c,d) show MuMax3 magnetisation simulations of a single ASVI vertex in all-macrospin (c) and wide-bar vortex, thin-bar macrospin (d) states with corresponding simulated magnetic force microscope (MFM) images in (e) and (f). Vortex bars exhibit a characteristic `checkered' pattern under MFM, with diagonally-opposite quadrants of positive (white) and negative (dark) magnetic charge. We note that the relative orientations of dark and light MFM-contrast quadrants allow vortex chirality readout (the opposite chirality magnetisation states in fig. \ref{Fig1}d) have opposite MFM contrast checkerboards in fig. \ref{Fig1}f). The reservoir computation scheme described below is unaffected by vortex chirality. Figure \ref{Fig1} g,h) show experimental MFM images of all-macrospin (g) and a vortex chain in an otherwise macrospin state (h). Vortices are slightly distorted in the experimental MFM image relative to simulation, due in part to tip-sample interactions favouring attractive (dark) over repulsive (light) interaction\cite{gartside2016novel}. A detailed study of the vortex formation process is provided in supplementary note 1 and supp. fig. 1.

To study how vorticisation progresses during field-cycling, Figure \ref{Fig1} i) shows an MFM image-series where an all-macrospin, $-\hat{x}$-saturated ASVI state (top-left panel) is subjected to four sequential $\pm$18 mT $\hat{x}$ minor field loops and imaged after each field. 18 mT is chosen such that thin bars never reverse while wide bars reverse each field application, save for those becoming pinned via local microstate-dependent dipolar-field textures (e.g. left and top edges of 3-loop, negative field panel). 18 mT is below the vortex-to-macrospin (V2M) conversion field (20 mT for the relative $\mathbf{H}_{\rm app}$ \& array orientation here), creating a ratchet effect where some macrospins convert to vortices each loop, but not vice-versa, increasing vortex population throughout field-cycling. V2M conversion is examined further in Figure \ref{Fig3}. For clarity, when describing `saturated' ASVI states we refer to the remanent state after applying a global magnetic field (typically $H_{sat} = 200~$ mT) such that all nanoislands are in a macrospin state, magnetised along the field axis e.g. top-left panel of fig. \ref{Fig1}i).

\begin{figure*}[thbp]   
\centering
\includegraphics[width=0.99\textwidth]{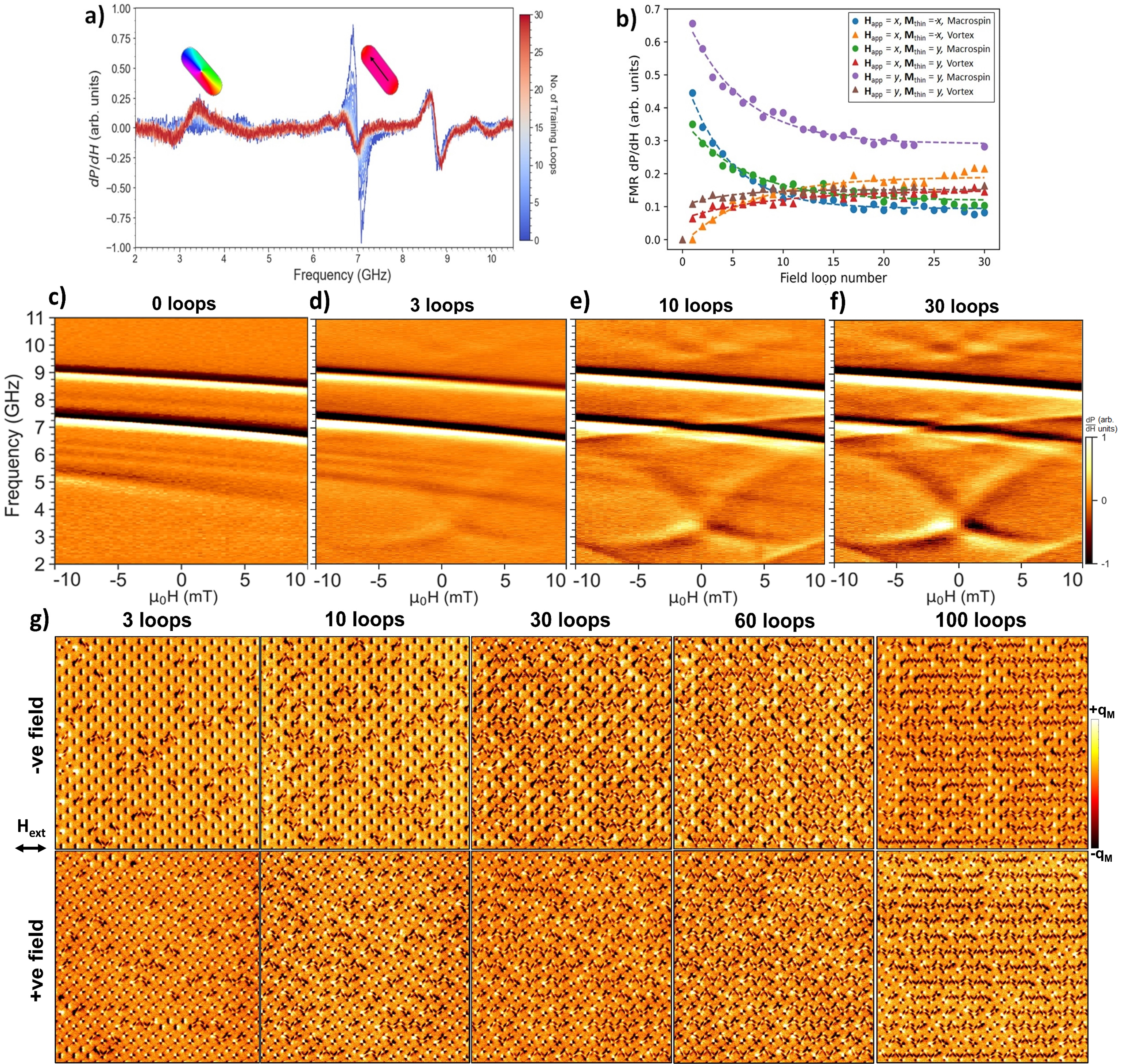}
\caption{Reconfigurably-directed vortex evolution and spin-wave spectra.\\ 
a) Differential FMR spectra measured in -1.2 mT bias field after 0-30 successive $\pm$18 mT field-cycles. Zero-loop state corresponds to -200 mT saturated pure-macrospin state. Four main modes are observed: low-frequency vortex mode ($\sim$3.5 GHz), wide-bar macrospin mode ($\sim$7 GHz), thin-bar macrospin mode ($\sim$8.8 GHz) and high-frequency vortex mode ($\sim$9.75 GHz).\\
b) Mode-amplitude of low-frequency vortex and wide-bar macrospin modes for 0-30 $\pm$18 mT field-cycles. Curves are displayed for three cases: 1 - Field along $\hat{x}$ orientation, $-\hat{x}$ saturated thin-bars (blue, orange points). 2 - Field along $\hat{x}$ orientation, $+\hat{y}$ saturated thin-bars (green, red points). 3 - Field along $\hat{y}$ orientation, $+\hat{y}$ saturated thin-bars (purple, brown points). Dashed lines for macrospin modes are exponential decay fits $y = Ae^{-\tau_{MS} x} + c$, with decay constant $\tau$ the vortex evolution rate and $c$ corresponding to the final macrospin population, vortex modes fits are $y = k - e^{-\tau_V x}$. Different vortex evolution rates and final vortex/macrospin populations are observed for the three cases, demonstrating reconfigurably-directed vortex evolution via thin-bar dipolar bias-field landscape and field $\hat{x}/\hat{y}$ orientation relative to the sample lattice.\\
c-f) Differential $\pm10$~mT FMR heatmaps measured after 0 (c), 3 (d), 10 (e) and 30 (f) $\pm$18 mT field cycles. Sample was initially saturated along $-\hat{x}$, field applied along $\hat{x}$. $\chi$-shaped low-frequency vortex mode and `checkerboard' high-frequency vortex modes increase with intensity throughout field-cycling.\\
g) MFM series of 3-100 $\pm$18 mT field cycle states imaged after negative (top row) and positive (bottom row) fields of each field-cycle. Increasing numbers of vortex-state bars are observed as field-cycling progresses, matching the increasing vortex FMR-mode intensity in c-f) heatmaps. Images are 15$\times15 \upmu$m$^2$.\\} 
\label{Fig2} \vspace{-1em}
\end{figure*}

Vortices initially appear with stochastic placement (Figure \ref{Fig1}k - 0 loops, positive (+ve) field). As field-cycling progresses, vorticisation occurs preferentially adjacent to existing vortices. This is due to the low dipolar field emanating from vortex bars causing asymmetry in the local dipolar-field texture and increasing likelihood of asymmetric field-torque on $Q_{\rm T} = +\frac{1}{2}$ defects during switching. This local promotion of vorticisation leads to formation of vortex and macrospin domains, with defined domain-structures taking shape by the 4th loop, +ve field image and clearly observed as field-cycling continues to 5-10 loops (supplementary figure 2 and note 2) and in higher loop-number images in Figure \ref{Fig2} g). A higher vorticisation probability is observed when moving from positive to negative field, $3.05\%~$ macrospins vorticising per loop vs. $1.34\%$ when switching from negative to positive. As mentioned above this is due to different microstates and dipolar-field landscapes between field polarities. Thin bars remain magnetised along $-\hat{x}$ while wide bars reverse, hence negative fields place macrospins in `type 2' spin-ice states\cite{gartside2021reconfigurable} (0 loop, -ve field panel) while positive fields give `type 1' or ground-states\cite{morgan2011thermal,nisoli2007ground} (macrospins in 0 loop, +ve field panel). The two states have differing dipolar-field landscapes, in type 2 wide and thin bars are magnetised the same way, giving symmetric dipolar field at the vertex, while type 1 has oppositely-magnetised wide and thin bars which gives an unbalanced dipolar-field texture due to the stronger dipolar-field of the wide bar. Again this is more likely to give unbalanced field-torques on $Q_{\rm T} = +\frac{1}{2}$ defects, driving them to combine to a $Q_{\rm T} = +1$ vortex state. 

To demonstrate vorticisation stochasticity, we compare three separate but identical field-cycling sequences, each beginning from saturated all-macrospin states. Figure \ref{Fig1} i), Figure \ref{Fig2} g) and supplementary figure 2 show different vortex locations and domain structures forming on the same array area, confirming vorticisation is a stochastically-dominated process, rather than determined by nanofabrication-imperfections termed `quenched disorder' which would favour spatially-similar domain patterns each field-cycling sequence. 

\section*{Reconfigurable spin-wave spectra and vortex state evolution}

We have observed via MFM with single-bar resolution how vorticisation occurs. MFM is an intrinsically slow process, each image takes 10-30 minutes with scan-windows limited $\sim10-100~$ \textmu m. It requires cumbersome mechanical apparatus, unsuitable for device integration. Ferromagnetic resonance (FMR) has emerged as a rapid, scaleable on-chip microstate readout technique well-suited to strongly-interacting nanomagnetic arrays\cite{vanstone2021spectral}. While not providing single-spin, exact microstate resolution, FMR can elucidate fine microstate details including ASI vertex-type populations and domain sizes\cite{vanstone2021spectral}, unavailable via magnetometry such as MOKE or VSM. Here we employ FMR to spectrally fingerprint mixed vortex-macrospin states. 

We analyse mode frequencies following the Kittel equation\cite{kittel1948theory} $f = \frac{\mu_0 \gamma}{2 \pi} \sqrt{\mathbf{H} \cdot (\mathbf{H}+\mathbf{M})}$ in the $k=0$ limit applicable to this work, $\gamma$ is the gyromagnetic ratio and $\mathbf{H} = \mathbf{H}_{\rm app} + \mathbf{H}_{\rm loc}$, the globally-applied field $\mathbf{H}_{\rm app}$ and the local dipolar-field of the nanomagnets $\mathbf{H}_{\rm loc}$. The local dipolar-field landscape varies greatly as vortex injection progress, with resulting distinct microstate-dependent magnon spectra. To focus on the effects of vortex injection on the microstate and vortex population, spectra are measured at a consistent small bias-field, chosen for good vortex-mode signal-to-noise. All spectral differences may therefore be attributed to microstate changes and corresponding shifts in $\mathbf{H}_{\rm loc}$\cite{vanstone2021spectral}. Broadband FMR spectra were measured in differential $\frac{\partial P}{\partial H}$ mode, 10 MHz frequency resolution with samples excited by mm-scale coplanar waveguide.

Figure \ref{Fig2} a) shows differential FMR spectra measured after the negative-field arm of each $\pm18~$ mT loop over a 30 loop field-cycling sequence. Initial 0-loop state (dark blue trace) is $-\hat{x}$ saturated, all-macrospin state, field is then applied along $\hat{x}$. Colour-scale denotes field-loop number, final 30-loop state is dark-red.
The initial all-macrospin state exhibits two modes, a wide-bar macrospin mode at 7 GHz and thin-bar macrospin mode at 8.8 GHz. As field-cycling progresses, the wide-bar mode decreases in amplitude as vortex-injection converts macrospins to vortices. Wide-bar macrospin mode-frequency redshifts throughout field-cycling as $\mathbf{H}_{\rm loc}$ is reduced by increasing numbers of flux-closed vortices, shifting 0.4 GHz after 30 loops. Similarly the thin-bar mode is blueshifted 0.15 GHz.
As the wide-bar macrospin mode decreases, a new 3.5 GHz vortex-mode grows, with equal vortex and macrospin mode amplitudes by 10 loops and vortex-mode amplitude double the macrospin at 30 loops. Fine shifts in mode amplitude and frequency are observed throughout field-cycling, demonstrating the capacity of vortex-injection to tailor relative mode power and frequency and provide on-demand spectral reconfiguration with more subtle, analogue-style control available than via reconfiguration of entire microstates\cite{gartside2021reconfigurable}. The correspondence of mode-amplitude to vortex and macrospin populations demonstrates the applicability of `spin-wave fingerprinting' to multi-texture spin-systems\cite{vanstone2021spectral}. The FMR response of ASI subjected to minor loops was previously studied in a conventional all-macrospin system\cite{jungfleisch2016dynamic} and is a powerful method for interrogating array microstate dynamics without slow MFM imaging or expensive beamline or LTEM techniques.

So far we have considered thin-bars as providing a static dipolar bias-field. We may exploit their magnetisation states as an extra degree of freedom and reconfigurably `direct' vortex injection. Figure \ref{Fig2} b) shows peak-amplitude extractions of wide-bar macrospin and vortex modes over 30-loop field-cycling sequences for three distinct cases: wide-bars and thin-bars initially saturated along $-\hat{x}$ as in Figure \ref{Fig2} a) (blue and orange traces), wide-bars saturated along $-\hat{x}$, thin-bars along $+\hat{y}$ (green and red traces), and wide and thin bars saturated along $+\hat{y}$ (brown and purple traces). Field is applied along wide-bar saturation axis in each case (i.e. $\hat{x}$, $\hat{x}$ and $\hat{y}$ respectively). Macrospin mode amplitude is fitted with $y = Ae^{-\tau_{MS} x} + c$, with decay constant $\tau_{MS}$ the macrospin mode-evolution rate and $c$ corresponding to the final macrospin population. Vortex mode amplitude is fitted as $y = k - Be^{-\tau_{V} x}$ with $\tau_V$ the vortex mode-evolution rate \& k relating to final vortex population. By mode-evolution we here refer to the changing mode power caused by the gradual conversion of the macrospin population to vortices by repeated field-cycling. 

For all cases, distinct mode-evolution rates and final vortex/macrospin populations are observed - showing the degree of control available from the thin bars and the sensitivity of vorticisation to dipolar-field texture. We can tailor vortex-injection dynamics via the reconfigurable bias-field from the thin bars. While parallel thin-bar states are prepared with uniform field here, one may locally define arbitrary thin-bar magnetisation states\cite{gartside2018realization,wang2016rewritable} to spatially texture vortex-injection.

Exploring vortex-mode field evolution, Figure \ref{Fig2} c-f) shows FMR heatmaps for 0-30 field-cycle states. For the 0-loop, all-macrospin state wide-bar ($\sim7~$ GHz) and thin-bar ($\sim9~$ GHz) nanobar-centre localised modes are observed, alongside a higher-index thin-bar mode ($\sim8~$ GHz) and three higher-index wide-bar macrospin modes ($\sim5,6,6.5~$ GHz). After 3 field cycles, new mode structures are observed with a pair of sigmoid-like modes between $2.5-6.5~$ GHz forming an $\chi$-shaped structure intersecting at $+1~$ mT. These striking new modes correspond to the vortex state, increasing in amplitude in the 10-loop heatmap as more bars vorticise. Checkerboard-pattern higher-index 9.5-10.5 GHz vortex modes with near-zero field-gradient also become visible at 10 loops. The 10-loop heatmap shows a lower-amplitude wide-bar macrospin mode with opposite (negative) gradient, corresponding to the population of oppositely-magnetised wide-bars (7.5 GHz mode at +10 mT), pinned by $\mathbf{H}_{\rm loc}$ as observed in higher loop-number MFM images. Vortex modes continue to increase in amplitude in the 30-loop heatmap, as does the oppositely-magnetised wide-bar macrospin mode. The opposing wide-bar macrospins cancel each other's dipolar-field, reducing net $\mathbf{H}_{\rm loc}$ and shifting the $\chi$-shaped vortex-mode intersection towards 0 mT. Vortex-mode field-gradients are highly non-linear, allowing enhanced control over mode-frequency in vortex-injected ASVI relative to conventional reconfigurable magnonic crystals and highlighting the degree of spectral-reconfigurability offered by ASVI. ASVI exhibits curved (low-frequency vortex modes), straight (macrospin modes) and flat (high-frequency vortex mode) mode gradients, an unusually rich spin-wave mode spectra for a nanopatterend reconfigurable magnonic system. Simulations also show a vortex-core gyrational mode\cite{chou2007direct,barman2010gyration} $\sim0.1-0.5~$ GHz, which we don't observe as our FMR is limited to 2 GHz minimum frequency.

Linking the spectral response to the microstate and showing the effects of extended field-cycling, Figure \ref{Fig2} g) shows MFM images at 3-100 field loops, with 3-30 loop states corresponding to the FMR heatmaps. The domain growth and increasing vortex population observed over 0-4 loops in Figure \ref{Fig1} i) continues, with defined domain patterns observed and high-purity vortex states reached by 100 loops.

\begin{figure*}[t]   
\centering
\includegraphics[width=0.9\textwidth]{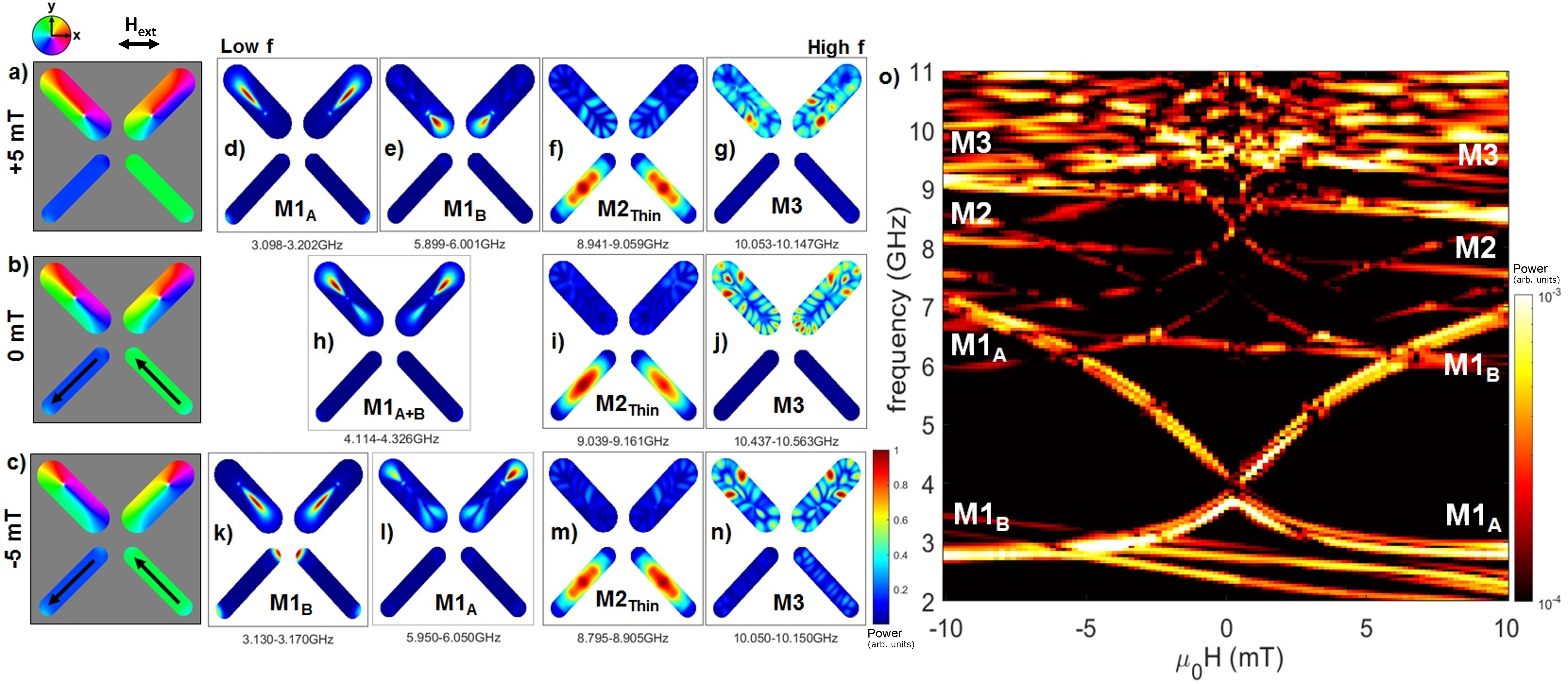}
\caption{Simulated spatial magnon mode-power maps and heatmap.\\
a-c) MuMax3 magnetisation states of ASVI vertex with both wide-bars in vortex state at $\mathbf{H}_{\rm app}$=+5 mT (a), 0 mT (b) and -5 mT (c). Vortex core is displaced along bar length by $\mathbf{H}_{\rm app}$, leading to two low-frequency 3-6 GHz modes corresponding to magnetisation regions above (M1$_A$) and below (M1$_B$) the vortex core. Both vortices here have the same chirality, M1$_A$ and M1$_B$ are inverted in terms of high/low frequency at a given field for opposite chirality vortices. \\
d-n) Spatial magnon mode power maps for the M1$_A$ and M1$_B$ vortex modes, M2 thin-bar macrospin mode and M3 whispering-gallery like vortex mode.\\
o) Simulated heatmap showing mode-dispersion with field.
} 
\label{Fig2B} \vspace{-1em}
\end{figure*}

\section*{Micromagnetic simulation of spin-wave spectra and spatial mode profiles}

Figure \ref{Fig2B} shows MuMax3 simulations of the spatial profiles of ASVI magnon modes at $\mathbf{H}_{\rm app}=$+5 mT (Figure \ref{Fig2B}a), 0 (b) and -5 (c) mT for a vertex with vortices in both wide bars. Three main modes are observed: M1$_A$ \& M1$_B$ are bulk-like modes localised above (A) and below (B) the vortex-core. M1 modes exhibit sigmoidal field gradients with opposite $\frac{\partial f}{\partial H}$-sign as $\mathbf{H}_{\rm app}$ causes the M1$_A$ region to grow at the expense of M1$_B$ as field is swept negative-to-positive and the vortex-core moves along the bar length. The macrospin thin-bars exhibit a bulk-mode, M2, and the vortex wide-bars exhibit a higher-order mode, M3, with whispering-gallery like profile around the bar edge\cite{schultheiss2019excitation}. Figure \ref{Fig2B} shows simulated mode field-evolution, showing good correspondence with experimental FMR minus the wide-bar macrospin mode, which is not present in the simulated vertex. Additional higher-index modes are resolved in simulation, these are more sensitive to nanopatterning imperfection and fall below experimental signal-to-noise threshold. The range of modes and their broad set of profiles and field-gradients are a strong example of the flexibility and benefits offered by magnetic-texture based magnonics\cite{yu2021magnetic}. 

\begin{figure*}[tbp]   
\centering
\includegraphics[width=0.97\textwidth]{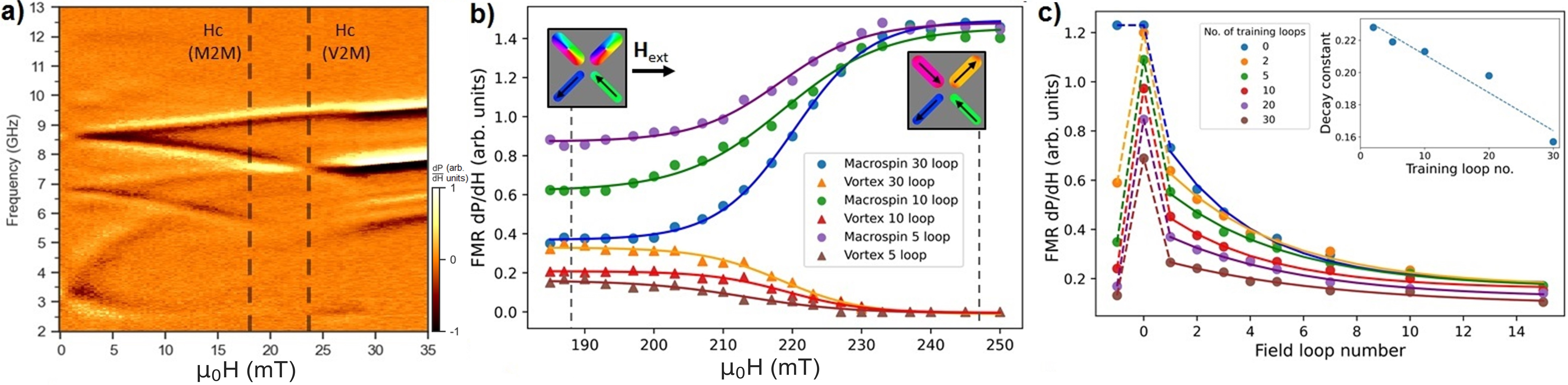}
\caption{Vortex-to-macrospin conversion and complex field-cycling sequences.\\
a) FMR heatmap showing 0-35 mT field sweep along $\hat{x}$ starting from high vortex-population state, ending in saturated pure-macrospin state. Initial high-vortex population state has thin-bars and macrospin wide-bars magnetised along $\hat{y}$. Wide-bar macrospins switch to $+\hat{x}$ magnetisation at 17 mT (Hc `M2M'), vortices switch to $+\hat{x}$ macrospins beginning at 24 mT (Hc `V2M') and saturate at 27 mT. \\
b) Mode-amplitudes of wide-bar vortex and macrospin modes while positively-sweeping field from 5, 10 and 30 $\pm$18 mT field-cycle states to saturated pure-macrospin state. Spectra were measured in -1.2 mT bias field after each field application to remove effects of varying $\mathbf{H}_{\rm app}$ on mode amplitudes, allowing clearer state comparison. Field was applied in $\hat{x}$ direction, thin bars saturated along $-\hat{x}$ hence V2M conversion occurs at slightly lower fields than panel (a). V2M conversion begins above 19.5 mT, allowing the ratchet effect whereby vorticisation is achieved with 18 mT field-cycles as vortices remain stable until higher $\mathbf{H}_{\rm app}$.\\
c) Macrospin FMR mode-amplitude response of 0,2,5,10,20 and 30 $\pm$18 mT field-cycle states to a single $+$21.5 mT `stimulus' field application, then 15 subsequent $\pm$18 mT `recovery' field-cycles. 
21.5 mT stimulus field is chosen following results of panel (b) to convert $\sim35-50\%$ of vortices to macrospins.
Initial field-cycled state amplitudes are shown at field loop number -1, 21.5 mT stimulus field is applied at loop 0, loops 1-15 are $\pm$18 mT field-cycles. States retain memory (i.e. longer-cycled states exhibit lower macrospin mode amplitudes) many loops after the 21.5 mT field application. Post-stimulus mode-evolution rate $\tau$ is fitted and plotted inset as a function of $n$ (number of pre-stimulus field-cycles). Linear fitting gives $\tau = -2.35\times 10^{-3}n + 0.235$, R$^2 = 0.94$. Field applied along $\hat{x}$.\\
} 
\label{Fig3} \vspace{-1em}
\end{figure*}

\section*{Vortex-to-macrospin conversion and fading memory behaviour}

The vorticisation process is bidirectional, with distinct switching dynamics and coercive fields when converting vortices to macrospins. Figure \ref{Fig3}a) shows an FMR heatmap starting at 0 mT with a 30-loop, high vortex-population state then sweeping $\mathbf{H}_{\rm app}$ 0-35 mT. Macrospins in the initial state are magnetised along $+\hat{y}$, $\mathbf{H}_{\rm app}$ swept along $+\hat{x}$. Three switching behaviours are observed: Wide-bar macrospins switching at 15.5-17 mT, thin-bar macrospins at 27 mT, and V2M conversion from 24-28 mT. The tapering linewidth of the $\sim7.5~$ GHz wide-bar macrospin mode between 24-28 mT reveals details of quenched-disorder effects in V2M-conversion (see supplementary note 3).

Figure \ref{Fig3} b) shows V2M conversion for 5, 10 and 30 field-cycle states, cycled at $\pm18~$ mT along $\hat{x}$ and swept 0-25 mT along $+\hat{x}$ until reaching a saturated all-macrospin state.
Phenomenological fits to macrospin and vortex mode-amplitudes are achieved using sigmoid functions. For all states V2M conversion begins at $\mathbf{H}_{\rm app} =~$19.5 mT, reaching saturated all-macrospin states at 23.8 mT. This gives a vortex-injection field window 18-19.5 mT above the wide-bar coercive field and below V2M conversion within which to exploit the vortex-injection ratchet effect. Thin-bar magnetisation can be used to control the V2M switching fields, with $-\hat{y}$ magnetised thin bars increasing V2M conversion to begin at 24 mT and reach an all-macrospin state at 28 mT. Thin-bar control over V2M conversion is discussed in detail in supplementary note 4 and supp. fig. 3.

Partial conversion of vortex populations to macrospins can affect subsequent microstate evolution behaviour. In Figure \ref{Fig3} c) we prepare 0-30 loop states at $\pm18~$ mT then apply a `stimulus' field of +21.5 mT, chosen to convert $\sim35-50\%$ of vortices to macrospins. After stimulus field application, the system is subjected to 15 $\pm18~$ mT $\hat{x}$ field-cycles. Wide-bar macrospin mode amplitude is measured at each step, with step number -1 the pre-stimulus state, step 0 the response to the stimulus field, and steps 1-15 the post-stimulus field-cycles.

Looking at initial pre-stimulus state amplitudes, as expected we observe lower macrospin amplitudes for longer field-cycles. After stimulus-field application (loop-number 0), we observe that the longer the field-cycling, the smaller the increase of macrospin mode-amplitude in response to stimulus field. This is somewhat surprising, as longer-cycled states have larger vortex populations, and therefore more bars available to convert to macrospins. This shows vortex-domains collectively resist V2M conversion. This is an important finding that shows ASVI retains long-term memory of its history, exhibiting substantially different responses to identical stimuli based on history.
We now examine response to post-stimulus $\pm18~$ mT loops. As in Figure \ref{Fig2} b), we observe exponential decay of macrospin mode-amplitude. We find the mode-evolution rate $\tau$ is a function of pre-stimulus field-cycling history $\tau = -2.35\times 10^{-3}n + 0.235$ with $n$ the number of pre-stimulus field-cycles, plotted inset in Figure \ref{Fig3} c). This shows the underlying vortex evolution dynamics are themselves history-dependent. Additionally, longer-cycled states exhibit lower macrospin mode amplitudes even 15 loops after the stimulus field. This shows ASVI vortex evolution history is a persistent and measurable property. One can distinguish shorter and longer preparation-cycle samples even after identical long measurement field-cycle sequences and stimulus-field applications. Crucially, while ASVI retains memory of its history, the different system trajectories in fig. \ref{Fig3} c) gradually converge at higher field-loop numbers. This progressive `forgetting' is termed fading memory, a key prerequisite for strong reservoir computing performance\cite{yildiz2012re,jensen2020reservoir}. Fading memory often occurs passively via intrinsic system damping\cite{torrejon2017neuromorphic,moon2019temporal}, here as in other schemes it is driven actively via the looping global field serving as a system clock\cite{jensen2020reservoir}. Longer sequences of varying magnitude field-loops (up to 250 loops) are examined in supplementary note 5 and supp. fig. 4, showing how field-magnitude changes as small as 0.5 mT exert substantial influence over microstate trajectory dynamics.

\begin{figure}[htbp]
\centering
\includegraphics[width = .92\textwidth]{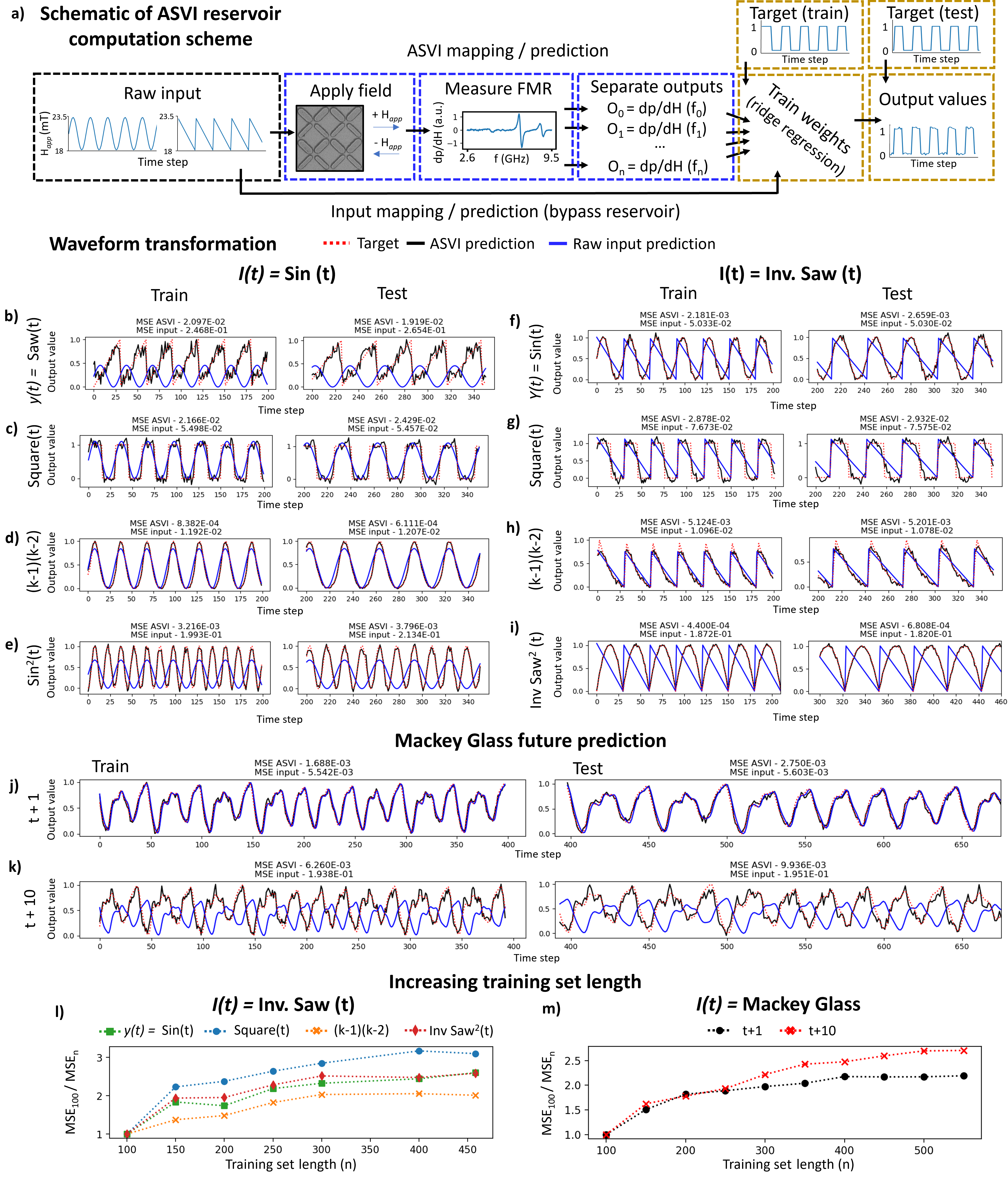}
\caption{ASVI reservoir time-series transformation and prediction.\\
a) Schematic of reservoir computing scheme. Input values 0-1 are scaled over $\mathbf{H}_{\rm app}=$18-23.5 mT field range. ASVI output response is obtained by applying a field loop then measuring FMR spectra at $\mathbf{H}_{\rm app}$ between 2.6 - 9.5 GHz (20 MHz steps). Weights are obtained via ridge-regression on the 'train' dataset and applied to a separate `test' dataset. 
b-i) Transformation of b-e) sine-wave and f-i) inverse saw-wave input datasets \textit{I(t)} to a variety of target waveforms \textit{y(t)}: b) saw-wave c,g) square-wave, d,h) second-order hysteretic non-linear transformation e,i) I\textsuperscript{2}(f\textsubscript{0}) and f) sin-wave. Here, we sample the reservoir outputs at 40 MHz from 3.6 - 4.56 GHz and 7.1 - 8.06 GHz (encompassing the vortex and macrospin peaks respectively) giving a total of 48 reservoir outputs.
Solid blue traces represent same training procedure performed on the raw \textit{I(t)} input datasets, bypassing the ASVI reservoir. Performance is severely reduced when bypassing the reservoir, visible as incorrect waveform shape and increased errors. \\
j,k) Future prediction of j) t+1 k) t+10 for a Mackey-Glass chaotic differential time-series, a commonly used reservoir computing benchmark. ASVI performs well while raw inputs fail to produce a true prediction, simply reproducing the input dataset with a t+1 time-lag j), failing badly at the more challenging t+10 task k). Here, all 397 reservoir outputs are used \\
Reservoir performance when lengthening the `train' dataset for l) saw-wave transformation and m) Mackey-Glass future prediction. Reservoir performance improves when increasing training length before plateauing.
}
\label{Fig5}
\end{figure}

\section*{Reservoir computation}

ASVI fulfills key RC criteria; its response to a given input is non-linear and history-dependent alongside its vast $4^N$ microstate-space\cite{jaeger2001echo,lukovsevivcius2009reservoir}. The crucial `fading-memory property'\cite{yildiz2012re,jensen2020reservoir} is also present in ASVI as demonstrated in Figure \ref{Fig3} c), where the response converges from different initial states when driven through the same data input field-loop sequence. 

Figure \ref{Fig5}  a) shows a schematic of the ASVI reservoir computing scheme. Reservoir computing schemes consist of three layers: an input layer, a `hidden' dynamic reservoir, and an output layer.
Here, the input-layer corresponds to globally-applied minor field loops with varying magnitude $\mathbf{H}_{\rm app}$, the hidden dynamic reservoir is the intrinsic nonlinearities (vortex-injection and spectral field response) and physical memory of the ASVI, and the output layer the measured FMR spectra after each input field-loop. 

Input values are linearly mapped onto an appropriate $\mathbf{H}_{\rm app}$ range (18.0-23.5 mT) such that thin bars never reverse. For each data-point, a minor field loop with maximum field of $\mathbf{H}_{\rm app}$ is applied along $\hat{x}$. The ASVI, serving as the reservoir, responds to the input-field in a non-linear fashion as shown throughout this work. The FMR response is then used as the output layer where spin-wave mode-amplitudes give microstate readout\cite{vanstone2021spectral}. We measure FMR between 2.6-10.5 GHz (20 MHz resolution) at $\mathbf{H}_{\rm app}$. The amplitude of each frequency bin is used as an independent output, giving a total of 397 reservoir outputs for each input time-step. We repeat this process for all inputs in the entire dataset before training, with prediction performed offline. 

The goal of the training is to assign a weight $w_i$ to each reservoir output $O_i$ such that \( \sum_{i=0}^{i=n} w_iO_i \sim Y \)  where $n$ is the number of reservoir outputs and $Y$ is the target value. This is performed on the entire training set so that the optimum set of weights is learned. For this, we use ridge regression via matrix multiplication (see Methods). When training, the measured dataset is split into `train' and `test' datasets. The `train' dataset is used to fit weights and the `test' dataset is used to assess fitted weights on previously unseen data. We employ short `train' datasets to mirror real-world, embedded device use-cases with strict limits on data capture, energy-cost and processing time. 
The weights are then applied to the `test' dataset which was not present during the weight learning, therefore assessing the computational performance on unseen data. We measure performance as the mean squared error (MSE) between the target waveform and the ASVI prediction. To remove noise and reduce over-fitting, we discard specific ASVI frequency-bins depending on task (see methods and supplementary note 6 and supp. fig. 5). For our training method, we consider only the outputs at a given time step - i.e. no time-multiplexing or software-based memory is employed, all memory effects occur physically within the ASVI reservoir. The advantage here is that no storage of past reservoir responses is required for inference after training is complete, reducing memory cost and processing time and avoiding deceptively-good reservoir performance where the software regression provides much of the computation. The training method employed here is simplistic by design, requiring no additional steps such as time-multiplexing or complex feedback which are costly in energy and processing time - vital considerations for embedded low-power, high-speed neuromorphic hardware\cite{nakajima2020physical,tanaka2019recent}.

We first assess the ASVI reservoir capacity to learn challenging non-linear transformations of a given input sequence \textit{I(t)} onto an unknown output function \textit{y(t)} = \textit{f(I(t))}. Figure \ref{Fig5}b-i) show results of learning non-linear transformations of sine-wave (b-e) [inverse-saw wave, (f-i)] input datasets \textit{I(t)} to \textit{y(t)} targets of: b) saw-wave [f) sine-wave] c) [g)] square-wave, d) [h)] a second-order non-linear hysteretic transformation following previous studies\cite{atiya2000new,du2017reservoir}   \(y(t) = 0.4y(t-1) + 0.4y(t-1)y(t-2) + 0.6I\textsuperscript{3}(t) + 0.1\) e) [i)] \textit{I\textsuperscript{2}(f\textsubscript{0})}. Here, we sample the reservoir outputs at 40 MHz from 3.6 - 4.56 GHz and 7.1 - 8.06 GHz (encompassing the vortex and macrospin peaks respectively) giving a total of 48 reservoir outputs. ASVI successfully learns to map all non-linear transformations of each input sequence, with strong performance competitive with existing RC schemes\cite{du2017reservoir} and low mean-square error (MSE) of 6.1 $\times$ 10\textsuperscript{-4} -  2.9 $\times$ 10\textsuperscript{-2} for the `test' dataset. 

The blue traces in Figure \ref{Fig5} represent the same training procedure applied to the raw input dataset, entirely bypassing the ASVI reservoir and showing only the effect of software regression - an important test for assessing RC performance. The raw input predictions are seen to entirely fail at the transformation tasks, their mapping attempts simply reproduce the input datasets with different amplitude scaling. ASVI succeeds here as the requisite non-linearities for strong mapping performance are supplied by the intrinsic physical nanomagnetic interactions and spectral field-response \cite{nakajima2020physical,tanaka2019recent}.

Next, we assess future prediction performance. We use the chaotic Mackey-Glass differential equation, commonly used for RC benchmarking\cite{moon2019temporal,tanaka2019recent,milano2021materia}. The same training method is used with the target the Mackey-Glass equation at t+$\tau$ where $\tau$ is how far into the future to predict. Here, all 397 reservoir outputs are used. Figure \ref{Fig5} j,k) show ASVI prediction for \textit{t+1} and \textit{t+10} with MSE values of 2.75 $\times$ 10\textsuperscript{-3} and 9.94 $\times$ 10\textsuperscript{-3} respectively. In this scenario, the raw input `prediction' is simply a reproduction of the input dataset with a $\tau$ time-step lag, a well-known behaviour indicative of prediction breakdown. Hence the ASVI prediction is superior despite the higher MSE, apparent when predicting further into the future as in the $\tau + 10$ case in Figure \ref{Fig5}k). The input prediction does not match the target value whereas the ASVI prediction closely follows the target trend. Further prediction steps are shown in supplementary note 7 and supp. fig. 6. 

Figure \ref{Fig5} l,m) show the reservoir performance when increasing the length of the `training set' for the inverse saw-wave transformation and Mackey-Glass prediction tasks respectively. Doing so allows for a truer ridge-regression fit, reducing the possibility of over-fitting. For both tasks we see improved reservoir performance up to 450-550 training points as demonstrated by the increasing ratio between MSE\textsubscript{100} / MSE\textsubscript{n} where n is the number of training points. Here, all 397 reservoir outputs are used.

We present a modified reservoir computation scheme in supplementary note 8 and supp. fig. 7 where all spectra are measured in a constant -1.2 mT bias field rather than at $\mathbf{H}_{\rm app}$. Computational performance is somewhat reduced with higher MSE values, though considerable performance is retained - demonstrating the versatility of ASVI across distinct measurement schemes.

It is worth examining the benefits offered by ASVI relative to other hardware reservoir computing systems. Nanomagnetic artificial spin systems have the benefit of no long-term degradation over repeated switching cycles, a substantial issue for current memristor-based systems which suffer from significant degradation and oxidation during repeated cycling. The WOx redox memristors used in recent reservoir computation studies\cite{moon2019temporal,du2017reservoir} have a maximum endurance of $10^{12}$ cycles before device breakdown\cite{wang2020resistive}. At a reasonable clock speed of 100 MHz (with far higher speeds desirable), this offers just 2 hrs 45 minutes of operation - a significant hurdle for the field to overcome. ASVI may be cycled indefinitely without degradation.
Moreover, nanomagnetic artificial spin systems such as ASVI have the benefit of indefinite non-volatile data storage and state retention, highly-attractive both for robustness against power supply issues (especially for ‘in-the-field’ applications) and long term retention of specific states with desirable functional characteristics. 
Additionally, there is no requirement to connect current-address lines to all individual nanomagnets within the array - beneficial to device cost, fabrication complexity, power-consumption and Ohmic heating. Nanomagnets intrinsically strongly-interact with their neighbours whereas such interactions must be artificially engineered into the system at-cost for memristors and other systems.

The scheme described here has substantial scope for future scalability and device-integrability via readily achievable modifications. Nanopatterned spin-wave resonators with high-Q at specific frequencies may be employed (e.g. at the macrospin and vortex peaks) for rapid readout at different spatial regions of micron-scale ASVI reservoirs. Field-input speeds for both data-input and RF-excitation can be dramatically increased to $10^7-10^8~$ Hz via ns-pulsed fields from nanopatterned striplines\cite{burn2017dynamic,pushp2013domain}, enabling continuous input of directly captured data.  Combining multiple nanopatterned striplines and pickup resonators or waveguides crucially allows for parallel as well as sequential data input and readout. 

ASVI is also well-matched to hybrid approaches combining reservoir computation with DNN-like manual tuning of network weights in hardware via direct nanomagnetic writing e.g. surface-probe\cite{gartside2016novel,gartside2018realization,gartside2020current,stenning2020magnonic,wang2016rewritable}, optically\cite{pancaldi2019selective,gypens2021thermoplasmonic,stenning2021low} or microwave-assisted switching\cite{bhat2020magnon}. A recent memristive network scheme by Milano et al has elegantly integrated network-weight optimisation into its hardware design\cite{milano2021materia}, eliminating the separate offline software training step. ASI has been shown to exhibit memristive functionality\cite{caravelli2019artificial,caravelli2021anisotropic} and implementing a scheme inspired by the Milano et al approach is a promising future avenue. As neuromorphic artificial spin systems mature, diverse technological applications including embedded medical implants, edge computing, remote sensors and smartphones stand to benefit from access to low-power neuromorphic computational hardware.

\section*{Conclusions}

We have demonstrated ASVI, a four-state spin-system with engineered texture bistability giving rise to emergent dynamics, including collective physical memory phenomena and highly-reconfigurable spin-wave spectra. Fading memory behaviour is observed when the system is driven through repeated minor field loops. Vortices exhibit substantial capacity to locally modify memory and switching behaviour, highlighting the rich emergent dynamics from engineering diverse magnetic textures in artificial spin systems. Vortex chains and domains may be harnessed to define magnon waveguides\cite{stenning2020magnonic,papp2021nanoscale}. The vortex-to-macrospin frequency shift of $\Delta f = 3.8~$ GHz is competitively high across reconfigurable magnonics and the analogue-style mode-amplitude tuning has technological appeal.

We demonstrate the efficacy of ASVI as a neuromorphic computation platform across a diverse range of tasks by learning linear and non-linear waveform transformations in addition to chaotic time-series prediction with strong results competitive with other reservoir systems\cite{du2017reservoir,tanaka2019recent,moon2019temporal} while employing short training datasets and no individual electrical-addressing of reservoir elements. We expect the scheme implemented here to be compatible across a range of artificial spin systems including conventional all-macrospin artificial spin ice, albeit without the benefits of clear frequency separation between macrospin and vortex peaks and asymmetric nonlinearities of the macrospin and vortex field-response. The observed large $\Delta f = 3.8~$ GHz mode shift far surpasses the 0.1-0.3 GHz shifts available in conventional all-macrospin ASI. Combining these benefits with the nanosecond timescales of the switching dynamics and spin-wave response invites  rapid, scaleable signal-processing and computation, expanding the scope of functional artificial spin systems. Developing low-energy hardware platforms for neuromorphic computation is a crucial effort as the energy-cost of machine learning rises exponentially. The low-power benefits of magnonics and intrinsic non-volatility, passive strong inter-element coupling and no current-address line requirement for each element situate ASVI as an attractive candidate for future neuromorphic computational hardware.

\subsection*{Acknowledgements}
WRB and JCG were supported by the Leverhulme Trust (RPG-2017-257 to WRB).\\
AV was supported by the EPSRC Centre for Doctoral Training in Advanced Characterisation of Materials (Grant No. EP/L015277/1).\\
TD was supported via International Research Fellow of Japan
Society for the Promotion of Science (Postdoctoral Fellowships
for Research in Japan).\\
The work of FC was carried out under the auspices of the NNSA of the U.S. DoE at LANL under Contract No. DE-AC52-06NA25396, and economic support of the LDRD grant PRD20190195.\\
Simulations were performed on the Imperial College London Research Computing Service\cite{hpc}.\\
The authors would like to thank Professor Lesley F. Cohen of Imperial College London for enlightening discussion and comments and David Mack for excellent laboratory management.
We would like to thank Ben Rogers, Karin Everschor-Sitte and Jake Love for valuable discussion regarding the reservoir computation scheme.

\subsection*{Author contributions}
JCG, KDS and AV conceived the work.\\
JCG drafted the manuscript other than the reservoir computation section, with contributions from all authors in editing and revision stages. KDS drafted the reservoir computation section with editing contributions from JCG.\\
JCG, KDS and AV performed FMR measurements.\\
JCG and HH performed MFM measurements.\\
JCG and KDS fabricated the ASVI. AV performed CAD design of the structures.\\
AV and JCG performed MOKE measurements of coercive field.\\
KDS implemented the reservoir computation scheme.\\
TD wrote code for simulation of the magnon spectra and performed micromagnetic simulations of mode dispersion relations and spatial mode profiles. TD performed mode character analysis and identification. \\
DMA wrote code for simulation of the magnon spectra. \\
FC provided valuable insight into the direction of the reservoir computation scheme. \\
HK contributed analysis of spin-wave dynamics. \\
WRB oversaw the project and provided critical feedback and direction throughout. \\

\subsection*{Competing interests}
The authors declare no competing interests.

\section*{Supplementary Information}

\section*{Methods}

Some description of our methodologies are reproduced from an earlier work of several of the authors\cite{gartside2021reconfigurable}, as the same methods are employed here.

\subsection*{Micromagnetic simulation}

Simulations were performed using MuMax3\cite{vansteenkiste2011mumax,vansteenkiste2014design}. To maintain field sweep history, ground state files are generated in a separate script and used as inputs for dynamic simulations. Material parameters for NiFe used are; saturation magnetisation, $M_{\rm sat}$ = 750 kA/m, exchange stiffness. $A_{\rm ex}$ = 13 pJ and damping, $\alpha$ = 0.001  All simulations are discretised with lateral dimensions, c$_{x,y}$ = 5 nm and normal direction, c$_z$ = 10 nm and periodic boundary conditions applied to generate lattice from unit cell. A broadband field excitation sinc pulse function is applied along z-direction with cutoff frequency = 20 GHz, amplitude = 0.5 mT. Simulation is run for 25 ns saving magnetisation every 25 ps. Static relaxed magnetisation at t = 0 is subtracted from all subsequent files to retain only dynamic components, which are then subject to an FFT along the time axis to generate a frequency spectrum. Power spectra across the field range are collated and plotted as a colour contour plot with resolution; $\Delta f$ = 40 MHz and $\Delta \mu_0 H$ = 1 mT. Spatial power maps are generated by integrating over a range determined by the full width half maximum of peak fits and plotting each cell as a pixel whose colour corresponds to its power. Each colour plot is normalised to the cell with highest power. High-resolution simulations performed for Figure 3 have lower damping, $\alpha = 0.0001$, and are run for 100 ns saving every 50 ps. The lower damping serves just to reduce linewidth for clarity of visualisation, and other behaviours associated with more realistic higher damping are well preserved\cite{stancil2009spin}. This produces colour plots with resolution; $\Delta f$ = 10 MHz and $\Delta \mu_0 H$ = 0.2 mT. $\mathbf{H}_{\rm app}$ is offset from the array $\hat{x},\hat{y}$-axes by $1^{\circ}$ to better match experiment\cite{gartside2021reconfigurable}. 

To achieve the vortex states (for example in fig. 1d) or fig. 3), MuMax3's built-in vortex initialisation command was used and then allowed to relax to a minimum energy configuration.

\subsection*{Nanofabrication}

ASVI was fabricated via electron-beam lithography liftoff method on a Raith eLine system with PMMA resist. Ni$_{81}$Fe$_{19}$ (permalloy) was thermally evaporated and capped with Al$_2$O$_3$. A `staircase' subset of bars was increased in width to reduce its coercive field relative to the thin subset, allowing independent subset reversal via global field.
The flip-chip FMR measurements require mm-scale nanostructure arrays. The 3x2 mm array employed for this study means that the distribution of nanofabrication imperfections termed `quenched disorder' is of greater magnitude here than typically observed in studies on smaller artificial spin systems, typically employing 10-100 micron-scale arrays. The chief consequence of this is that the Gaussian spread of coercive fields is over a few mT for each bar subset (15.5-17 mT for wide bars, 26-29 mT for thin)\cite{gartside2021reconfigurable}. Smaller ASVI arrays have narrower coercive field distributions, with the only consequence being that optimal applied field ranges
for reservoir computation input will be scaled across a corresponding narrower field range, not an issue for typical 0.1 mT or better field resolution of modern magnet systems. 

\subsection*{FMR measurement}

Ferromagnetic resonance spectra were measured using a NanOsc Instruments cryoFMR in a Quantum Design Physical Properties Measurement System. Broadband FMR measurements were carried out on large area samples $(\sim 2 \times 2~ \text{ mm}^2)$ mounted flip-chip style on a coplanar waveguide. The waveguide was connected to a microwave generator, coupling RF magnetic fields to the sample. The output from waveguide was rectified using an RF-diode detector. Measurements were done in fixed in-plane field while the RF frequency was swept in 10 MHz steps. The DC field was then modulated at 490 Hz with a 0.48 mT RMS field and the diode voltage response measured via lock-in. The experimental spectra show the derivative output of the microwave signal as a function of field and frequency. The normalised differential spectra are displayed as false-colour images with symmetric log colour scale\cite{gartside2021reconfigurable}.

\subsection*{Reservoir computation}

The reservoir training inputs are chosen to have approximately 30 data points per period for the sin,  inverse-saw waves and Mackey-Glass input with outputs sampled at every input. The Mackey-Glass time-delay differential equation takes the form $(\frac{dx}{dt} = \beta \frac{x_{\tau}}{1+x_{\tau}^{n}} - \lambda x)$ and is evaluated numerically with $\beta$ = 0.2, n = 10 and $\tau$ = 17. The array is initially saturated in a -200 mT field in the $\hat{x}$ direction. 

Reservoir computing schemes consist of three layers: an input layer, a 'hidden' reservoir layer, and an outputs layer corresponding the globally-applied fields, the ASVI and the FMR response respectively.

In each case, the inputs were linearly mapped to a field range spanning 18-23.5 mT, with the mapped field value corresponding to the maximum field of a minor loop applied to the system. After each minor loop, the FMR response is measured at the applied field $\mathbf{H}_{\rm app}$ and -1.2 mT between 2.6 - 10.5 GHz in 20 MHz steps. We perform measurements at two fields to compare the reservoir prediction quality. The FMR output is smoothed in frequency by applying a low-pass filter to reduce noise (examples of prediction quality without smoothing are shown in supplementary note 4). For each input data-point, we measure 397 distinct frequency bins and take each bin as an output giving 397 reservoir outputs. This process is repeated for the entire dataset with training and prediction performed offline. 

Offline training and prediction is performed using a matrix multiplication method. We first separate the ASVI response into two datasets. A `train' dataset for learning the optimum set of weights for a given task and a `test' dataset to test the performance of the learned weights on previously unseen data. If we consider the training set of reservoir outputs  \(\vec x_{\rm train}(u)\) and the target waveform \(\tilde y\)

\begin{equation*}
\vec x_{\rm train}(u) = 
\begin{pmatrix}
O_{0}(0) & O_{0}(1) & \cdots & O_{0}(m) \\
O_{1}(0) & O_{1}(1) & \cdots & O_{1}(m) \\
\vdots  & \vdots  & \ddots & \vdots  \\
O_{n}(0) & O_{n}(1) & \cdots & O_{n}(m) 
\end{pmatrix}
\\
\tilde y = 
\begin{pmatrix}
y\textsubscript{0} \\
y\textsubscript{1} \\
\vdots   \\
y_{n} 
\end{pmatrix}
\end{equation*}
where $m$ is the number of reservoir outputs, $n$ is the number of input data points in the `train' dataset and \(O_n(m)\) is the m\textsuperscript{th} reservoir output for input step n (i.e. the amplitude of the m\textsuperscript{th} frequency bin). The goal of training is to obtain a $1 \times M$ array of weights \(W_{\rm out}\) which transform the reservoir outputs  (\(\vec x(u)\)) into the desired target (\(\tilde y\)). For this, we use ridge regression which solves the linear optimisation problem  \(W\textsubscript{out}= \rm{argmin}_W (||\tilde y-W \vec x_{\rm train}(u)||^2+\lambda ||W||^2)\) with the following solution \(W\textsubscript{out} = (\vec x_{\rm train}(u)\textsuperscript{T }\vec x_{\rm train}(u) + \lambda I)\textsuperscript{-1} \vec x_{\rm train}(u)\textsuperscript{T}\tilde y \) where \(I\) is the identity matrix and $\lambda$ is the ridge regularisation term (fixed at 0.0001 for all tasks). This can be performed with freely-available Python packages such as scikit-learn.

Once \(W_{\rm out}\) is obtained, we can then multiply the `test' dataset \(\vec x_{\rm test}(u)\) with \(W_{\rm out}\) to obtain the reservoir prediction \(\tilde y_{\rm ASVI}\) (i.e. \(\vec x_{\rm test}(u) W_{\rm out} = \tilde y_{\rm ASVI} \ \)) as follows:

\[
\vec x_{\rm test}(u) W_{\rm out} = 
\begin{pmatrix}
O_{n+1}(0) & O_{n+1}(1) & \cdots & O_{n+1}(m) \\
O_{n+2}(0) & O_{n+2}(1) & \cdots & O_{n+2}(m) \\
\vdots  & \vdots  & \ddots & \vdots  \\
O_{n+l}(0) & O_{n+l}(1) & \cdots & O_{n+l}(m) 
\end{pmatrix}
\begin{pmatrix}
W\textsubscript{0} \\
W\textsubscript{1} \\
\vdots   \\
W_{m} 
\end{pmatrix}
 = \begin{pmatrix}
y\textsubscript{ASVI,\;\textit{n+1}} \\
y\textsubscript{ASVI,\;\textit{n+2}} \\
\vdots   \\
y\textsubscript{ASVI,\;\textit{n+l}} 
\end{pmatrix}
 = \tilde y_{\rm ASVI}
\]
where $l$ is the number of input datapoints in the `test' dataset. We assess the performance of the reservoir by calculating the mean-squared error (MSE) of target waveform and the reservoir response \(MSE = (\tilde y_{\rm ASVI} - \tilde y)^2/l \). 

To reduce noise and overfitting we chose a subset of the total outputs depending on the task. For the transformation tasks we sample the output at 40 MHz from 3.6 - 4.56 GHz and 7.1 - 8.06 GHz (encompassing the vortex and macrospin peaks respectively) giving a total of 48 reservoir outputs. For the Mackey-Glass future prediction we use all 397 reservoir outputs. For all transformation tasks except the inv. saw$^2$ task (Figure 5 b-h)) we use `train' and `test' lengths of 250 and 150 data points respectively. For the inverse saw$^2$ (Figure 5 i)) task we use a `train' length of 300 and a `test' length of 250. This is necessary due to the difficulty of the task. For the Mackey-Glass future prediction (Figure 5 j,k)) we use `train' and `test' lengths of 393 and 271 respectively. When exploring the effects of varying the `train' length (Figure 5 l,m)) we use a fixed `test' length of 100 data points. See supplementary note 4 for details on output selection. While the vortex mode is lower in amplitude than the macrospin mode, it is broader in frequency and therefore registered by more frequency bins which allows good computation performance.

\subsection*{MFM measurement}

Magnetic force micrographs were produced on a Dimension 3100 using commercially available normal-moment MFM tips\cite{gartside2021reconfigurable}.

\subsection*{Data availability statement}
The datasets generated during and/or analysed during the current study are available from the corresponding author on reasonable request.

\subsection*{Code availability statement}
The code used in this study is available from the corresponding author on reasonable request.

\subsection*{Supplementary note 1 - Detailed discussion of micromagnetic macrospin-to-vortex conversion process}


Supplementary figure 1 a) shows a simulated MuMax3 time series of the vorticisation process. The magnetisation texture of the bottom-right bar is distorted via a combination of applied field $\mathbf{H}_{\rm app} =$~16 mT and local dipolar field $\mathbf{H}_{\rm loc}$ from the other bars, resulting in vortex core formation and stabilisation on a nanosecond timescale. 

The detailed dynamics of the process can be well understood through a topological-defect picture, which we will now examine. Excellent references for describing magnetisation textures such as macrospins, vortices and domain walls via dynamic topological defects can be found in works by Pushp et al\cite{pushp2013domain} and Tchernyshyov et al\cite{tchernyshyov2005fractional}.

An initial $-\hat{x}$ saturated vertex (t=0) is field-swept along $+\hat{x}$ with a 1$^{\circ}$ angular-offset such that the bottom-right wide-bar experiences slightly higher field-torque and switches first. $\mathbf{H}_{\rm app}$ brings the topological charge $Q_{\rm T} = +\frac{1}{2}$ edge-bound topological defects at either bar-end to opposite long-edges (t=0.4~ns), creating pockets of $+\hat{x}$ magnetisation (red regions) which spread through the bar (t=0.58~ns). In normal macrospin reversal, the $\mathbf{M} = +\hat{x}$ region growth continues and the $Q_{\rm T} = +\frac{1}{2}$ defects finish at opposite bar ends. However, in vorticisation one of the $Q_{\rm T} = +\frac{1}{2}$ defects reverses direction halfway (t=0.78~ns). This is the crucial step differentiating macrospin-to-macrospin reversal from macrospin-to-vortex conversion. The defects now come into close proximity at the vertex-centre bar end (t=1~ns) before combining into a single $Q_{\rm T} = +1$-defect (t=1.25~ns). Integer-charge defects may only exist in the magnet bulk, and a $Q_{\rm T} = +1$-defect is otherwise known as a vortex-core. The vortex-core moves into the nanomagnet bulk (t=1.75~ns) before reaching a central equilibrium point, minimising exchange and demagnetisation energy (t=2.43~ns). The factors causing one $Q_{\rm T} = +\frac{1}{2}$ defect to reverse direction and drive vorticisation may be isolated in simulation, but are more stochastic in experiment. Angularly offsetting $\mathbf{H}_{\rm app}$ from $\hat{x}$ or $\hat{y}$ encourages vorticisation in simulation by generating unequal field torques on the $Q_{\rm T} = +\frac{1}{2}$ defects at either macrospin end. This effect was not observed in experiment, possibly due to edge-roughness affecting edge-defect trajectories and stochastic room-temperature thermal effects versus effective 0 K simulation. Simulation and experiment both find vorticisation more common when beginning from a type-1 microstate due to unbalanced dipolar fields from opposingly-magnetised thin and wide bars generating unequal field-torque on the two $Q_{\rm T} = +\frac{1}{2}$ defects.


Supplementary figure 1 b) shows a MuMax3 times series of V2M conversion at $\mathbf{H}_{\rm app} =$~21 mT. Vortex-cores are pushed by $\mathbf{H}_{\rm app}$ towards bar edges (t=0.48 ns) where they decompose on contact (t=0.68-0.88 ns) into pairs of edge-bound $+\frac{1}{2}$-defects characterising the macrospin state (t=1.5 ns). The edge-bound defect pair are visible in the t=0.68 ns frame, and are circled in black for clarity. Bringing the vortex core into proximity to the nanoisland edge carries a considerable exchange energy penalty\cite{tchernyshyov2005fractional} which must be overcome by $\mathbf{H}_{\rm app}$. Such a penalty is not present when switching between macrospin states, hence the higher coercive fields when switching out of a vortex state.

\begin{figure*}[thbp]   
\centering
\includegraphics[width=0.6\textwidth]{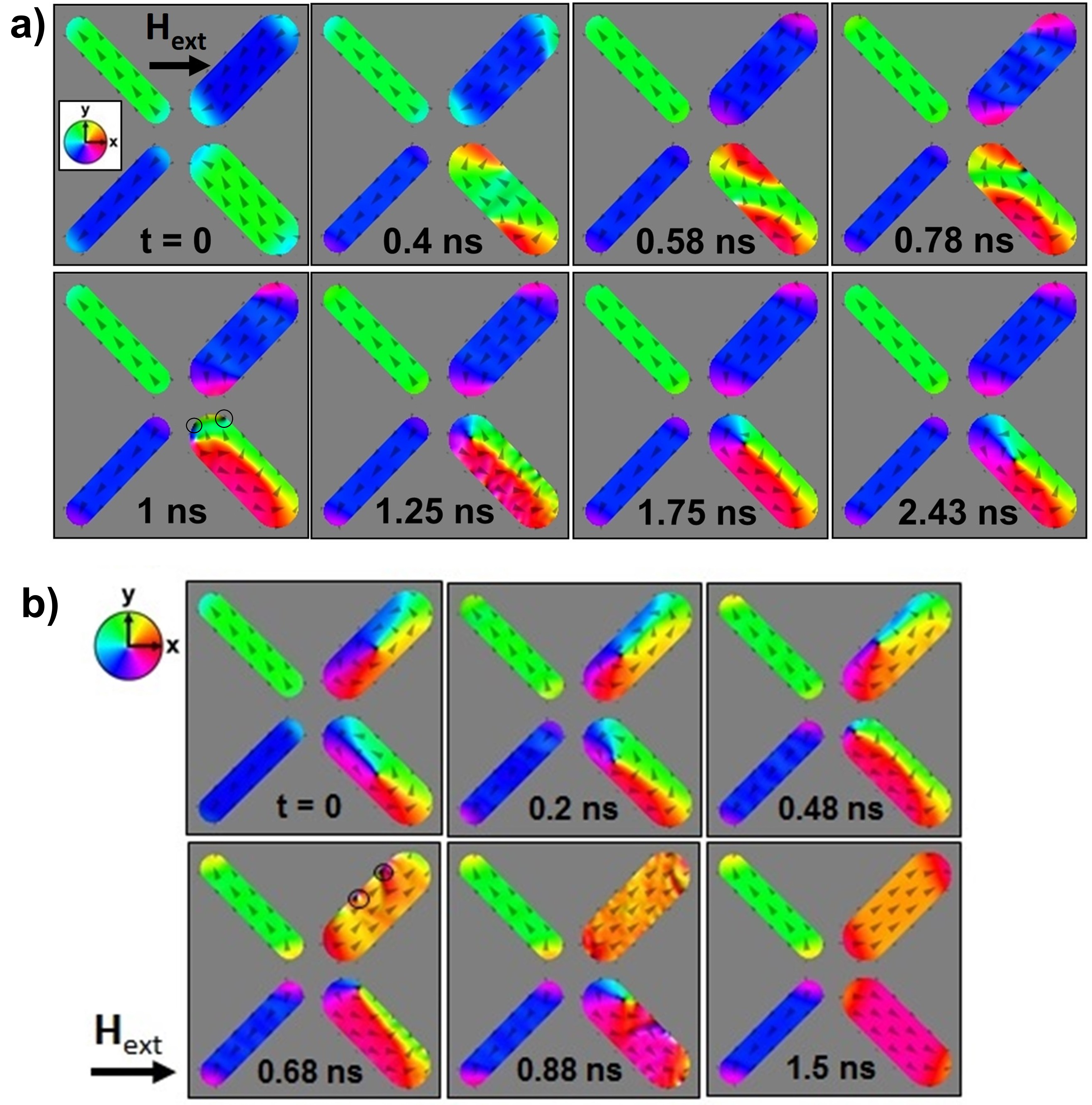}
\caption*{\textbf{Supplementary Figure 1}\\
a) MuMax3 time-series of vorticisation process at $\mathbf{H}_{\rm app}=16~$ mT. Macrospin state in bottom right bar changes to a vortex via topological-defect exchange combining two half-integer, edge bound defects from opposite bar ends in the macrospin state into a single $+1$~ winding number topological vortex-core defect in the bulk of the nanomagnet. Edge-bound half-integer $+\frac{1}{2}$ topological defects are circled in black in the 1 ns panel for visibility. \\
b) MuMax3 simulated time series of vortex-to-macrospin conversion at $\mathbf{H}_{\rm app}$ = 21 mT showing vortex-cores pushed by $\mathbf{H}_{\rm app}$ into nanoisland edges, decomposing vortices into macrospin states. The vortex cores decompose on contact with the nanowire edges to a pair of $+\frac{1}{2}$ topological defects, which are circled in black in the 0.68 ns panel for visibility. Field applied along $\hat{x}$.
}
\label{Suppfig1}
\end{figure*}

\subsection*{Supplementary note 2 - Extended 5-10 loop training sequence and distinct domain structures of distinct training sequences}

\begin{figure*}[thbp]   
\centering
\includegraphics[width=0.99\textwidth]{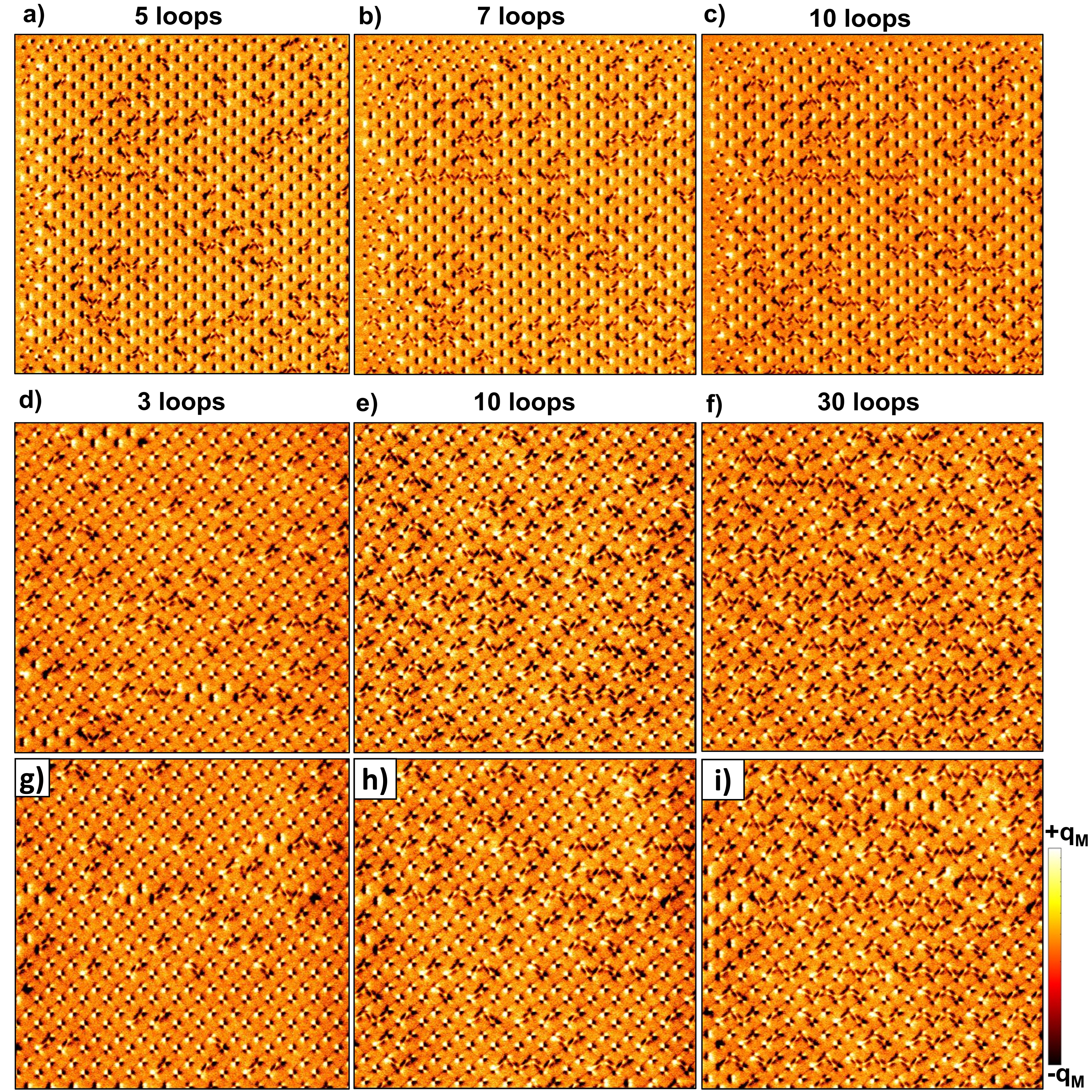}
\caption*{\textbf{Supplementary Figure 2}\\
a-c) 5-10 loop training sequence continuing from the 0-4 loop sequence in Figure 1 i).\\
d-f) 3-30 loop training sequence, imaged after positive field arm.\\
g-i) Subsequent 3-30 loop training sequence on the same array area as d-f), sample is reset to all-macrospin state with 200 mT field after sequence d-f). Different vortex locations and domain patterns are observed, highlighting the stochastic rather than deterministic nature of vortex training.
}
\label{Supp1}
\end{figure*}

Supplementary Figure 2 a-c) follow on from Figure 1 i), showing progressively higher vortex population as training progresses and more defined distinct vortex and macrospin domains, along with `trapped' macrospins along the top and left edges which remain pinned and do not reverse with each loop.

Supplementary Figure 2 d-f) and g-i) show two separate 3-30 loop training sequences on the same array area, with the sample reset (saturated to an all-macrospin state) between the two sequences. Locations of vortex bars and the spatial domain patterns are different in each training sequence, demonstrating that vortex training is stochastically-dominated, rather than a repeating process with the same bars vorticising every time determined by quenched disorder.

\subsection*{Supplementary note 3 - Quenched disorder influence on vortex-to-macrospin conversion and spin-wave mode linewidth and frequency}

Interesting details of V2M switching are observed by following the central frequency of the wide-bar macrospin mode (border between light and dark bands) as switching progresses from 24-28 mT. If we extend the Kittel mode gradient of the saturated wide-bar macrospin mode back from its linear region 28-35 mT, the central mode frequency diverges from this gradient between 24-28 mT. Measured mode frequency is higher than the expected Kittel gradient at 24 mT, and gradually decreases while linewidth increases until it meets the Kittel gradient at 28 mT. This is due to the Gaussian distribution of bar widths and corresponding resonant frequency spread caused by quenched disorder. While wider macrospin bars with lower resonant frequency tend to reverse at lower field\cite{vanstone2021spectral}, here we are switching from vortex to macrospin states rather than between oppositely-magnetised macrospins. Figure 1 b) shows thinner bars energetically favour macrospin states, and as such switch to macrospins at lower field than broader bars. Thinner bars also exhibit higher-frequency resonances, so as V2M conversion progresses the macrospin population is initially dominated by thinner, higher-frequency bars and as such shifts the central mode-frequency above the expected average resonant Kittel frequency.

\subsection*{Supplementary note 4 - Influence of thin bar magnetisation and dipolar bias-field on vortex-to-macrospin conversion process}

\begin{figure*}[thbp]   
\centering
\includegraphics[width=0.7\textwidth]{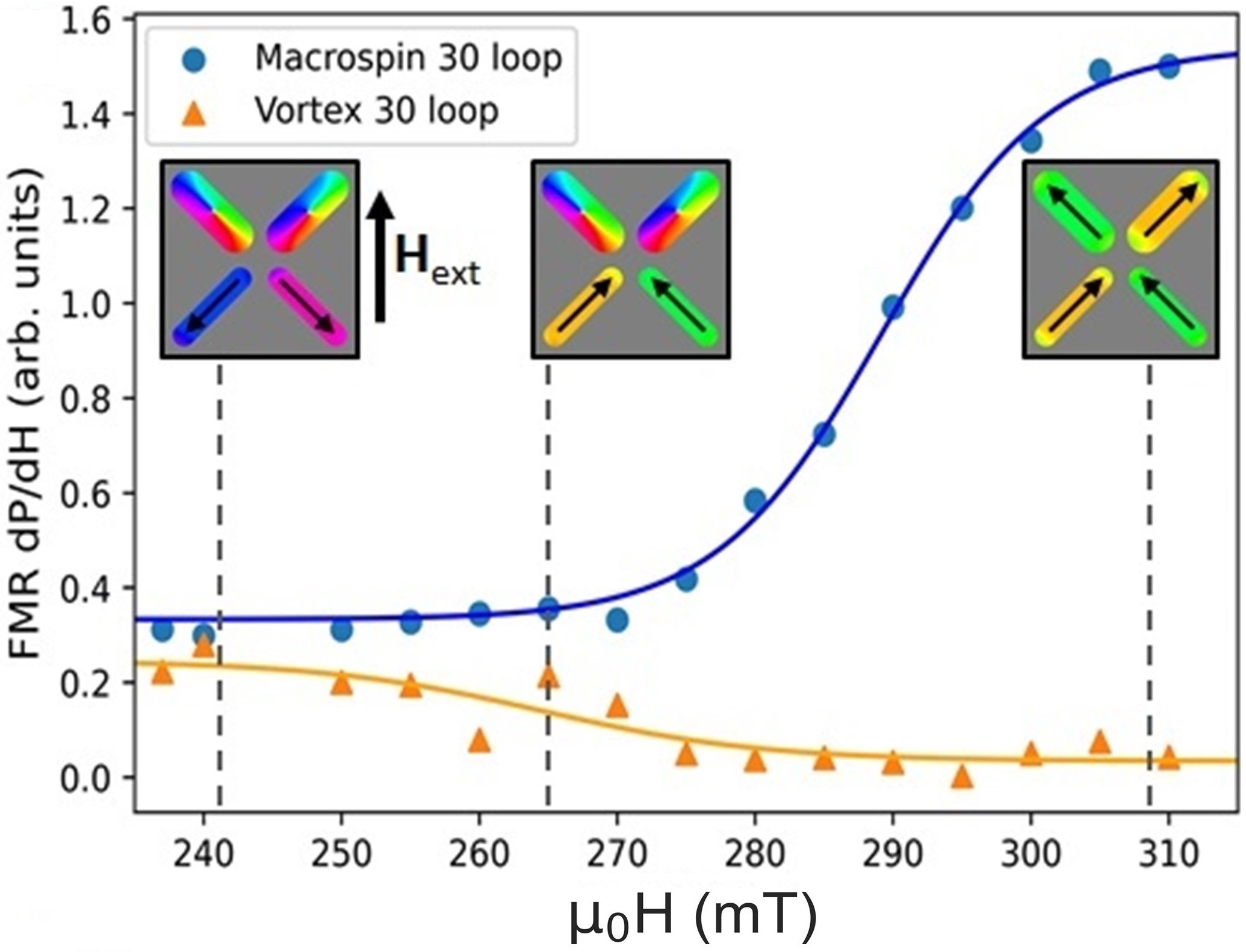}
\caption*{\textbf{Supplementary Figure 3}\\
Mode-amplitudes of wide-bar macrospin and vortex modes while positively increasing field along $+\hat{y}$ after 30 $\pm$19 mT $\hat{y}$ field-cycles, ending in saturated pure-macrospin state. Sample initially saturated along $-\hat{y}$ pre-cycling, thin bars initially saturated along $-\hat{y}$ before reversing at 27 mT and unlocking V2M conversion. Spectra measured in -1.2 mT bias field after each field application. V2M conversion begins at 27 mT, significantly higher than in fig. 4 (b). This demonstrates control over protected vortex field-range via reconfigurable thin-bar bias-field.\\
}
\label{Suppfig3}
\end{figure*}

Supplementary figure 3 demonstrates further the reconfigurable control provided by the thin-bars. We prepare a 30-loop state with $\mathbf{H}_{\rm app}=\pm19~$ mT along $\hat{y}$ with $-\hat{y}$ magnetised thin bars then sweep 0-31 mT along $+\hat{y}$ until saturating into macrospin state. Here, V2M conversion is prevented until a higher field, beginning at 26.5 mT and saturating at 30.5 mT (versus 19.5-23.8 mT for the case of thin-bars magnetised along $-\hat{x}$ as in fig. 4 b). This is due to dipolar vertex energetics. With $\mathbf{H}_{\rm app}$ along $\hat{y}$ and $-\hat{y}~$ saturated thin bars, vortices converting to macrospins would enter the `type 4' or `monopole' state\cite{ladak2010direct,gartside2021reconfigurable}, highly energetically-unfavourable repulsive configurations which impede motion of vortex-cores towards bar-edges. 

\subsection*{Supplementary note 5 - System response and microstate dynamics for longer sequences of varying-magnitude field-loops}

\begin{figure*}[thbp]   
\centering
\includegraphics[width=0.7\textwidth]{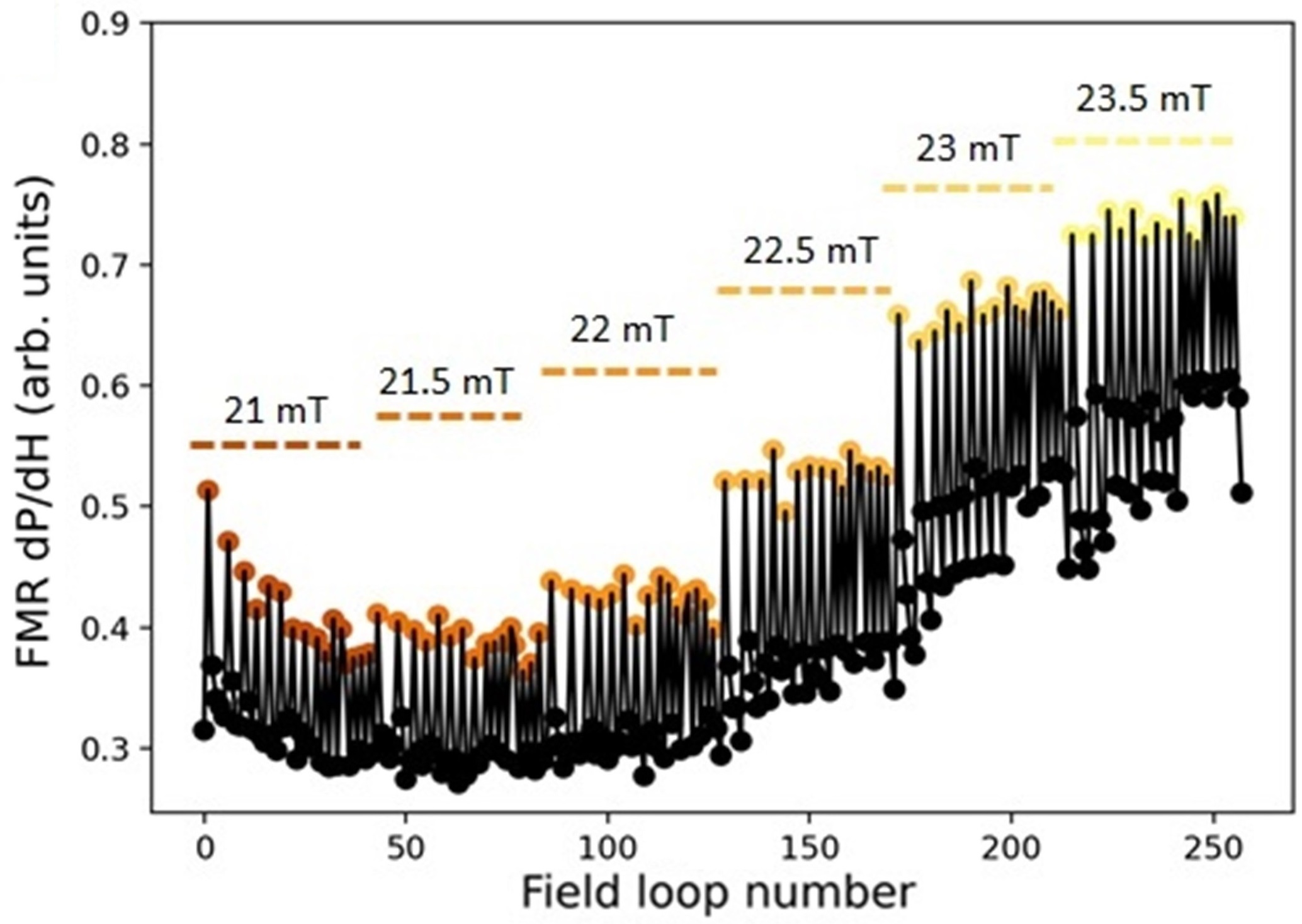}
\caption*{\textbf{Supplementary Figure 4}\\
Macrospin mode-amplitude evolution over field-cycle series consisting of single 21-23.5 mT `stimulus' field applications followed by two subsequent $\pm18~$ mT loops. Stimulus fields convert vortices to macrospins, 18 mT loops convert macrospins to vortices. Sensitive response of mode-amplitude gradient to applied-field magnitude is observed. System begins at loop 0 in a highly-vorticised state. \\
}
\label{Suppfig4}
\end{figure*}

ASVI is highly-sensitive to small changes in applied-field amplitude, leading to complex nonlinear responses to field-cycle sequences comprising different field-amplitudes. Supplementary figure 4 shows macrospin mode-amplitude evolution over a field sequence comprising single 21-23.5 mT `stimulus' field applications followed by two subsequent $\pm18~$ mT loops. System begins at loop 0 in a highly-vorticised state. Stimulus fields convert vortices to macrospins, 18 mT loops convert macrospins to vortices. Stimulus-field amplitude changes of 0.5 mT are enough to modify microstate evolution behaviour and mode-amplitude gradient, shown by the diverse range of amplitude gradients over the field sequence.

\subsection*{Supplementary note 6 - Reservoir output selection and effects of smoothing}

\begin{figure*}[thbp]   
\centering
\includegraphics[width=0.99\textwidth]{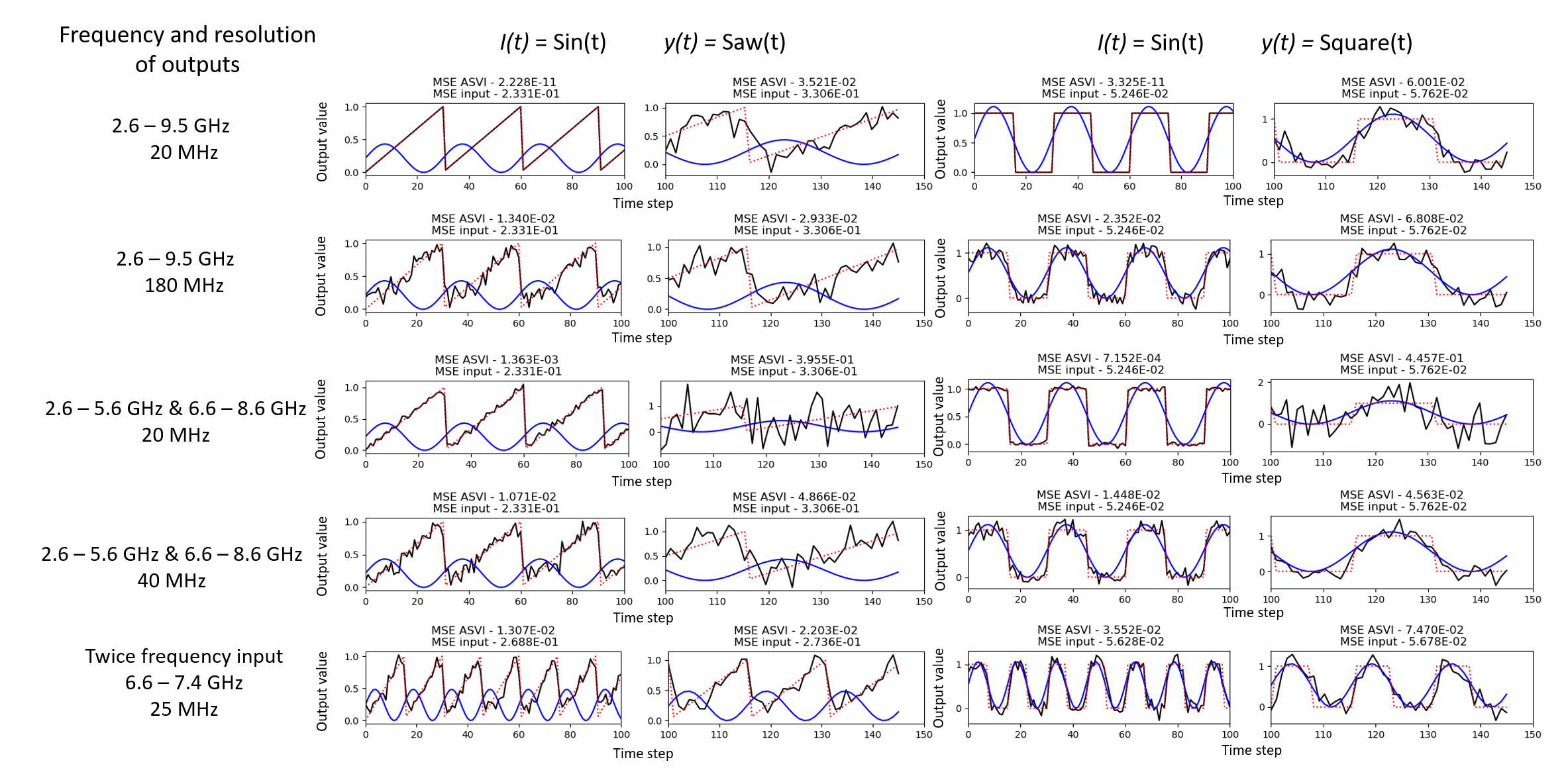}
\caption*{\textbf{Supplementary Figure 5}\\
Comparison between various output selections for reconstructing and saw and square wave from a sine-wave input. The optimum output selection is task dependent. Here, we do not smooth the FMR response before prediction.
}
\label{Suppfig5}
\end{figure*}

Supplementary Figure 5 shows a comparison for a variety of different output options when learning to map a sine-wave to a saw-wave and a square-wave. Here, we measure the FMR response at -1.2 mT and the FMR output does not undergo any smoothing.  The optimum output configuration, demonstrated by the lowest `test' MSE, is task dependent. Furthermore, the sampling rate of the output also affects the prediction performance. When comparing to Figure 5, reasonable prediction is still obtained in the absence of FMR smoothing. 

\subsection*{Supplementary note 7 - Mackey-Glass future predictions}

\begin{figure*}[thbp]   
\centering
\includegraphics[width=0.99\textwidth]{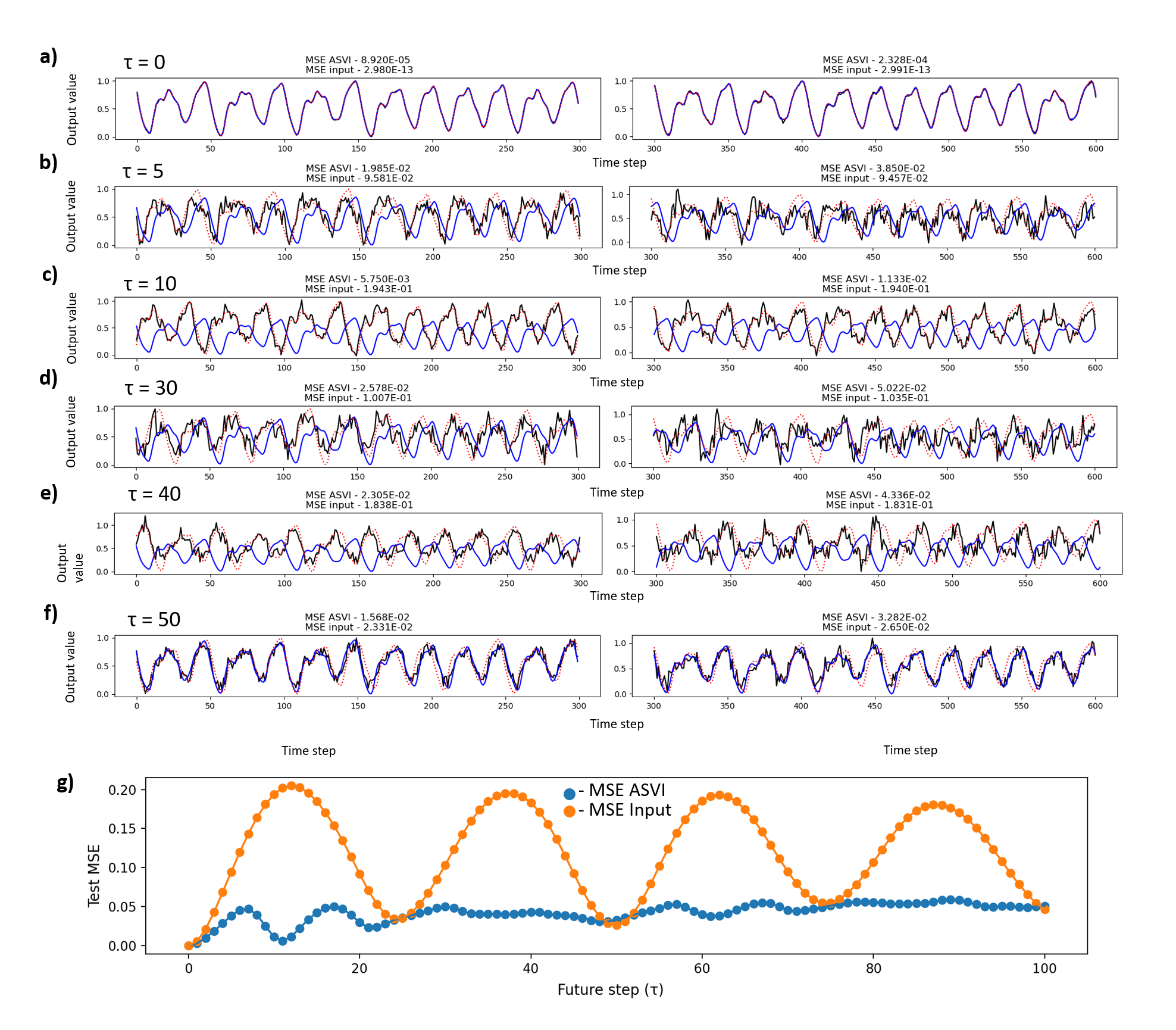}
\caption*{\textbf{Supplementary Figure 6}\\
 `Train' and `test' response for $\tau$ = a) 0, b) 5, c) 10, d) 30, e) 40 and f) 50 when forecasting the Mackey-Glass equation. g) `Test' MSE for Mackey-Glass prediction for up to 100 future steps. The ASVI response outperforms the input prediction when the target waveform is dissimilar to the input waveform. This is due to the input prediction simply reproducing the target dataset with a $\tau$ time-step lag rather than a true prediction of future performance - resulting in a sinusoidal MSE profile with peaks corresponding to $\pi$/2 phase shifts (e.g. panel e) $\tau$ = 40) and troughs corresponding to $\pi$ phase shifts (e.g. panel f) $\tau$ = 50)
}
\label{Suppfig6}
\end{figure*}

Supplementary Figure 6 shows the results for Mackey-Glass predictions from 0 - 100 points into the future. Here, we measure the FMR response at $H_{ \rm app}$ and ASVI outputs are taken as the FMR amplitude from 2.6 - 10.5 GHz at 20 MHz steps giving a total of 397 outputs. The ASVI prediction outperforms the input prediction when the target and input waveforms are dissimilar. While the Mackey-Glass differential equation is chaotic, it has an underlying approximate periodicity (here around 22-25 time steps). This leads to deceptively low MSE for the input `prediction' at integer multiples of this period, as it is in-fact simply a reproduction of the target dataset with a $\tau$ time-step lag, a well-known behaviour indicative of a breakdown in prediction performance. This results in a sinusoidal error profile and is therefore not a meaningful prediction. The peaks of the error profile correspond to a $\pi$/2 phase shift (panel e) $\tau$ = 40) and the troughs corresponds to a $\pi$ phase shift (panel f) $\tau$ = 50).  In each case, the inputs were linearly mapped to a field range spanning 18-23.5 mT, with the mapped field value corresponding to the maximum field of a minor loop applied to the system. The ASVI also displays initial oscillations in prediction performance, but significantly reduced relative to the software-only, raw input prediction. 

\subsection*{Supplementary note 8 - Measuring reservoir output spectra at constant -1.2 mT bias field}
Supplementary Figure 7 shows ASVI waveform transformation and Mackey-Glass future prediction when measuring the FMR spectra at a -1.2 mT bias field. Reasonable transformation is observed throughout demonstrating the versatility of ASVI across distinct measurement schemes

Supplementary Tables 1 $\&$ 2 show the MSE values and ratios when measuring at $H_{\rm app}$ and -1.2 mT for waveform transformation and Mackey-Glass future prediction respectively, MSE values range from 1.39 - 54.3 $\times$ lower when measuring at $H_{\rm app}$ due to the additional non-linear field-dependent mode shifts when measuring at the applied field. 

\begin{figure*}[htbp]
\centering
\includegraphics[width = .95\textwidth]{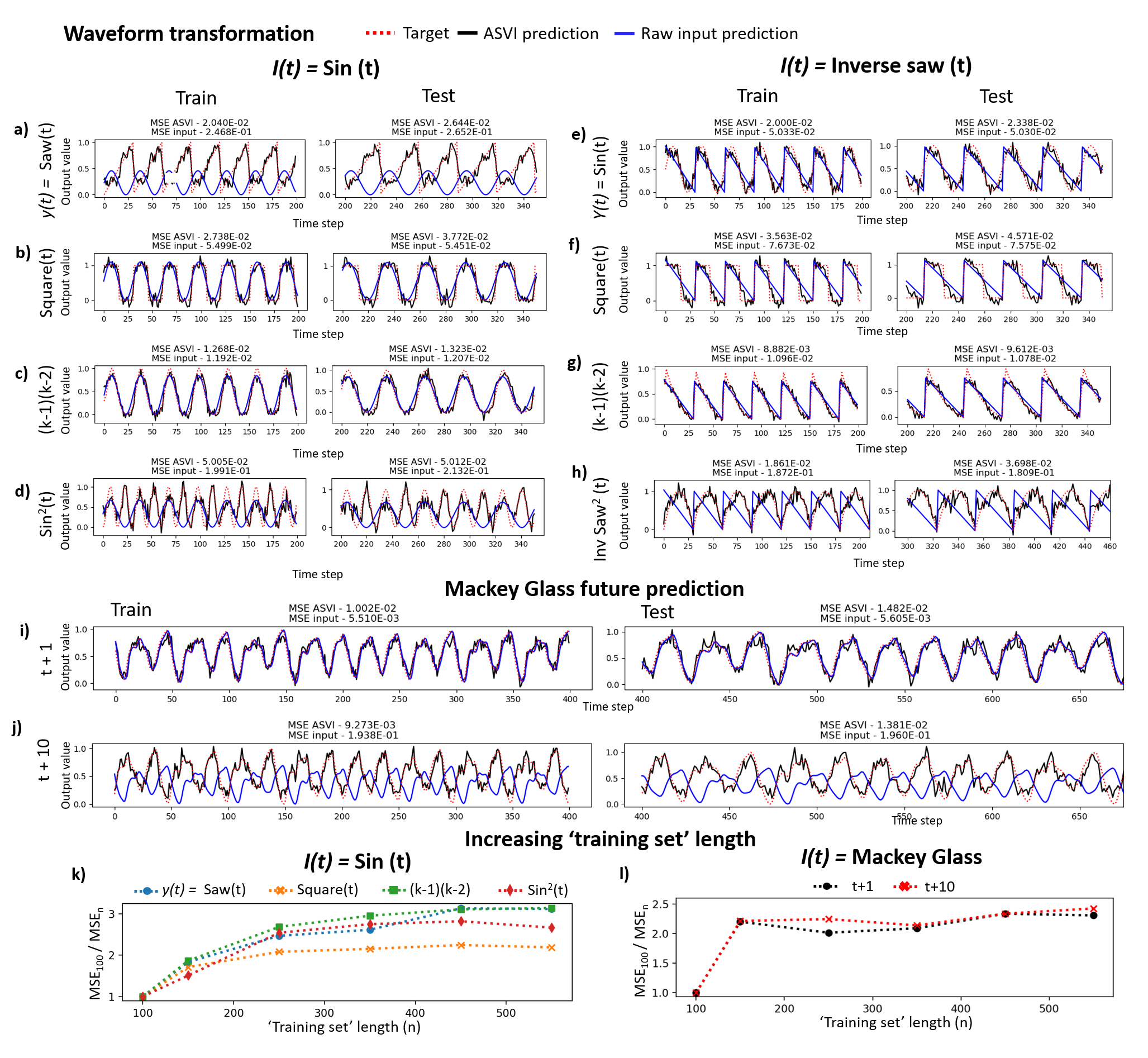}
\caption*{\textbf{Supplementary Figure 7}\\ ASVI reservoir time-series transformation and prediction measuring at -1.2 mT.\\
a-h) Transformation of a-d) sine-wave and e-h) inverse saw-wave input datasets \textit{I(t)} to a variety of target waveforms \textit{y(t)}: a) saw-wave b,f) square-wave, c,g) second-order hysteretic non-linear transformation d,h) I\textsuperscript{2}(f\textsubscript{0}) e) sin-wave. The ASVI reservoir maintains reasonable transformation performance when measuring at a static -1.2 mT bias field to a high degree of accuracy demonstrated by low `test' MSE errors (MSE ASVI).
Solid blue traces are the same training procedure performed on the raw \textit{I(t)} input datasets, bypassing the ASVI reservoir. Performance is severely reduced when bypassing the reservoir, visible as incorrect waveform curvature and increased errors.\\
i,j) Prediction of i) t+1 j) t+10 for a Mackey-Glass time-differential delay input commonly used as a benchmark for reservoir prediction quality. ASVI performs well while raw inputs fail to produce a true prediction and simply reproduce the input dataset with a t+1 time-lag j), which breaks down at the more challenging t+10 task k) resulting in poor performance. Reservoir performance when increasing the size of the `train' dataset for k) sine-wave transformation and l) Mackey-Glass future prediction tasks. In all cases, the reservoir performance improves when increasing the number of training points as seen by the increasing ratio of MSE\textsubscript{100}/MSE\textsubscript{n} where n is the length of the `training set'.
}
\label{Supp7}
\end{figure*}

\begin{table}[ht!]
\begin{tabular}{l|lll|lll}

\textbf{Task/Input (measurement field)} &\textbf{ Sin ($H_{\rm app}$)} & \textbf{Sin ($-1.2 mT$)} & \textbf{Improv. ($\times$)} & \textbf{Saw ($H_{\rm app}$)} & Saw \textbf{($-1.2 mT$)} & \textbf{Improv.  ($\times$)} \\ \hline
Sin / Saw                      & 1.92E-02   & 2.66E-02  & 1.39    & 2.66E-03  & 2.34E-02    & 8.79    \\ 
Square                         & 2.43E-02   & 3.77E-02  & 1.55    & 2.93E-02  & 4.57E-02    & 1.56   \\ 
(k-1)(k-2)                     & 6.11E-04   & 1.32E-02  & 21.6    & 5.20E-03  & 9.61E-02    & 18.5    \\ 
I$^2$         & 3.80E-03   & 5.01E-02  & 13.2    & 6.81E-04  & 3.70E-02    & 54.3    \\ 
\end{tabular}
\caption{\textbf{Supplementary table 1}\\
Comparison of MSE values for the waveform transformation tasks when measuring the FMR spectra at the $H_{\rm app}$ and -1.2 mT. MSE values are between 1.39 - 54.3 $\times$ lower when measuring at $H_{\rm app}$ due to additional non-linear field-dependent shifts in mode frequency.
}
\end{table}
\begin{table}[h!]
\begin{tabular}{l|lll}
\textbf{Task/Input (measurement field)} &\textbf{ MG ($H_{\rm app}$)} & \textbf{MG ($-1.2 mT$)} & \textbf{Improv. $\times$} \\ \hline
t + 1                          & 2.75E-03  & 1.48E-02    & 5.39   \\
t + 10                         & 9.94E-03  & 1.38E-02    & 1.39   \\ 
\end{tabular}
\caption{\textbf{Supplementary table 2}\\
Comparison of MSE values for Mackey-Glass prediction when measuring the FMR spectra at the $H_{\rm app}$ and -1.2 mT. MSE values are 5.39 and 1.39 $\times$ lower for t+1 and t+10 predictions respectively when measuring at $H_{\rm app}$ due to additional non-linear field-dependent shifts in mode frequency.}
\end{table}

\bibliography{Bibliography.bib}

\end{document}